\documentclass[%
 aip,
 amsmath,amssymb,
preprint,%
]{revtex4-1}

\usepackage{graphicx}
\usepackage{dcolumn}
\usepackage{bm}
\usepackage{xcolor}

\usepackage[utf8]{inputenc}
\usepackage[T1]{fontenc}
\usepackage{mathptmx}
\usepackage{subfigure}

\begin{document}

\preprint{AIP/123-QED}

\title[How motifs condition critical thresholds: Linking Micro- to Macro-scales]{How motifs condition critical thresholds for tipping cascades in complex networks: Linking Micro- to Macro-scales}

\author{Nico Wunderling}
\altaffiliation{These authors equally contributed to this study. Correspondences should be addressed to: nico.wunderling@pik-potsdam.de}
\affiliation{ 
Earth System Analysis, Potsdam Institute for Climate Impact Research (PIK), Member of the Leibniz Association, 14473 Potsdam, Germany
}
\affiliation{ 
Institute of Physics and Astronomy, University of Potsdam, 14476 Potsdam, Germany
}
\affiliation{ 
Department of Physics, Humboldt University of Berlin, 12489 Berlin, Germany
}

\author{Benedikt Stumpf}%
\altaffiliation{These authors equally contributed to this study. Correspondences should be addressed to: nico.wunderling@pik-potsdam.de}
\affiliation{ 
Earth System Analysis, Potsdam Institute for Climate Impact Research (PIK), Member of the Leibniz Association, 14473 Potsdam, Germany
}
\affiliation{ 
Department of Physics, Free University of Berlin, 14195 Berlin, Germany
}

\author{Jonathan Kr\"onke}
\affiliation{ 
Earth System Analysis, Potsdam Institute for Climate Impact Research (PIK), Member of the Leibniz Association, 14473 Potsdam, Germany
}
\affiliation{ 
Institute of Physics and Astronomy, University of Potsdam, 14476 Potsdam, Germany
}

\author{Arie Staal}
\affiliation{ 
Stockholm Resilience Centre, Stockholm University, Stockholm, SE-10691, Sweden
}

\author{Obbe A. Tuinenburg}
\affiliation{ 
Stockholm Resilience Centre, Stockholm University, Stockholm, SE-10691, Sweden
}
\affiliation{ 
Copernicus Institute, Faculty of Geosciences, Utrecht University, 3584 CB Utrecht, Netherlands
}

\author{Ricarda Winkelmann}
\affiliation{ 
Earth System Analysis, Potsdam Institute for Climate Impact Research (PIK), Member of the Leibniz Association, 14473 Potsdam, Germany
}
\affiliation{ 
Institute of Physics and Astronomy, University of Potsdam, 14476 Potsdam, Germany
}

\author{Jonathan F. Donges}
\affiliation{ 
Earth System Analysis, Potsdam Institute for Climate Impact Research (PIK), Member of the Leibniz Association, 14473 Potsdam, Germany
}
\affiliation{ 
Stockholm Resilience Centre, Stockholm University, Stockholm, SE-10691, Sweden
}

\date{\today}

\begin{abstract}
In this study, we investigate how specific micro interaction structures (motifs) affect the occurrence of tipping cascades on networks of stylized tipping elements. We compare the properties of cascades in Erd\H{o}s-Rényi networks and an exemplary moisture recycling network of the Amazon rainforest. Within these networks, decisive small-scale motifs are the feed forward loop, the secondary feed forward loop, the zero loop and the neighboring loop.\\
Of all motifs, the feed forward loop motif stands out in tipping cascades since it decreases the critical coupling strength necessary to initiate a cascade more than the other motifs. We find that for this motif, the reduction of critical coupling strength is 11\% less than the critical coupling of a pair of tipping elements. For highly connected networks, our analysis reveals that coupled feed forward loops coincide with a strong 90\% decrease of the critical coupling strength.\\
For the highly clustered moisture recycling network in the Amazon, we observe regions of very high motif occurrence for each of the four investigated motifs suggesting that these regions are more vulnerable. The occurrence of motifs is found to be one order of magnitude higher than in a random Erd\H{o}s-Rényi network.\\
This emphasizes the importance of local interaction structures for the emergence of global cascades and the stability of the network as a whole.  
\end{abstract}
                                                                                                                                                                                                                             
\maketitle

\begin{quotation}
Tipping elements are nonlinear systems, where a small perturbation can be sufficient to induce a qualitative change of the whole system as soon as a critical threshold (tipping point) is crossed. Coupled tipping elements exist for instance in connected lake systems, in the Earth's climate system or in social systems. Here, we investigate networks of interacting tipping elements, where each node consists of a stylized tipping element and explore important interaction structures on the micro scale of the network, the so-called \textit{motifs}. Such motifs in complex networks have been found in multiple systems such as cell metabolism, food webs or neural networks and are known to be significantly overexpressed in real-world compared to random networks. However, motifs have not yet been studied extensively in complex networks, where nodes have their own dynamics. In our study, we find that tipping cascades occur more often at locations with high motif frequency revealing locations (nodes) of decreased robustness. 
\end{quotation}

\section{\label{sec:level1}Introduction}
Methodologies from complex networks science have gained increasing attention since they have successfully been applied to a broad range of different fields ranging from physical sciences, biology or ecology to information transfer, energy systems and sociology~\citep{newman2003structure}. In many cases, network nodes are reasonably represented by continuous, nonlinear dynamical systems as, for instance, in oscillators in power grids, population dynamics in food webs or synchronization of nonlinear oscillators~\citep{zou2013reviving,gross2009generalized,nitzbon2017deciphering}. More recently, one focus of research shifted to the investigation of interacting tipping elements. Tipping elements are systems in which a small perturbation can lead to a qualitative change in the system in case a critical value (\textit{tipping point}) is surpassed. Tipping elements have been identified in the Earth's climate system~\citep{Lenton2008}, but also in various other contexts like finance, politics, ecology or climate~\citep{Brummit2015, Kriegler2009, Cai2016, rocha2018cascading}. In the Earth system, tipping elements can interact across scales in time and space~\citep{rocha2018cascading,gaucherel2017potential, dekker2018cascading} which could in principle lead to feedbacks, domino effects~\citep{klose2019dynamic} and ultimately to a hothouse state~\citep{steffen2018trajectories}.\\
Lately, these two approaches, complex networks and tipping elements, have been linked together in a conceptual approach to study cascading failure on networks~\citep{kronke2019dynamics, eom2018resilience}. Here, each node of such a network is a tipping element and has its own dynamics compared to other studies where cascading failure has been studied with discrete states of network nodes and a fixed threshold beyond which failure of the respective node is induced~\citep{watts2002simple,buldyrev2010catastrophic}. The links of the network then consist of any arbitrary positive or negative coupling, potentially with different weights, between the network nodes. This procedure yields a set of connected differential equations that can be described well by a network approach. If the network nodes are indeed tipping elements, the occurrence of tipping cascades, the failure of at least two nodes together, can be investigated. The dependence of cascades based on different interaction structures resembling the structure of paradigmatic network types like Erd\H{o}s-Rényi, small-world or scale-free networks has been assessed~\citep{kronke2019dynamics}. However, as we find here, in a certain regime of coupling strengths between the nodes, the dynamics of the whole network are dominated by local structures within the network. These sub-structures are the so-called \textit{motifs}.\\
Contrasting other recent publications reflecting the influence of the general network topology of cascading failures in complex networks~\citep{turalska2019cascading,loppini2019critical,wu2018degree,liu2019multiple} and how local interaction patterns determine the dynamics of their larger parent networks~\citep{krishnagopal2017synchronization,d2008synchronization,gambuzza2016amplitude}, this work aims to reveal how these local, small-scale structures condition tipping cascades within the whole system.\\
The notion of motifs has been introduced by Milo et al.~\citep{Milo2002} as the basic building blocks of complex networks. It has been shown that motifs can be identified for instance in food webs~\citep{stouffer2012evolutionary}, authorship attribution~\citep{marinho2016authorship} up to transcriptional networks that control the expression of genes~\citep{alon2007network}, e.g., in tumor suppressors or E. Coli~\citep{lahav2004dynamics, anastasiadou2018non, shen2002network, mangan2006incoherent}. The so-called \textit{feed forward loop} is an essential motif in such networks since it is significantly overexpressed in these real-world networks compared to typical random graphs~\citep{Milo2002}. Furthermore, the feed forward loop motif has been used to identify functionally important nodes in various real-world networks through the aggregation of several such motifs into clusters. This has been investigated among others in transcription networks of E. Coli, online Wikipedia networks or air-traffic~\citep{gorochowski2018organization} and hints at a special role of this motif as it efficiently passes system dependent information forward.\\
Here, we examine how selected micro-structures within an Erd\H{o}s-Rényi network of tipping elements significantly alter the occurrence of tipping cascades and with that the stability of the whole network (Fig.~\ref{fig:one}). We investigate these features on Erd\H{o}s-Rényi networks since their properties are controllable and reproducible. Furthermore, we look at the scaling behavior of motif occurrence and we are able to predict critical couplings in dense Erd\H{o}s-Rényi networks which can be traced back to coupled feed forward loops. Additionally, we compare our results for this to a real-world example, the moisture recycling network structure of the Amazon rainforest and point out important differences.

\section{Methods}
\subsection{System of differential equations}
In this study, the dynamics of each of the nodes in the network follows the autonomous ordinary differential equation
\begin{equation}
    \frac{dx}{dt}=-a(x-x_0)^3+b(x-x_0)+c,
\label{eq:cusp}
\end{equation}
when interactions are ignored. Here, $c$ is the critical individual forcing parameter, $a,b>0$ and $x_0$ represents a shift on the x-axis~\citep{kronke2019dynamics,klose2019dynamic}. This equation is unistable below a certain critical parameter $c_\text{crit,\ low}$ and above $c_\text{crit,\ high}$. In between, the system is bistable and state transitions occur via a saddle-node bifurcation at $c_\text{crit,\ low}$ and $c_\text{crit,\ high}$. Equation~\ref{eq:cusp} is a minimal example for continuous dynamical systems that possess two distinct stable states. Hence, this model can act as a paradigmatic model and has been applied to ecosystems like shallow lakes, but also ice sheets or the thermohaline circulation~\citep{Brummit2015,van2007theory,scheffer2007regime,scheffer2001catastrophic}. The bifurcation diagram of one of these tipping elements is shown in Fig.~\ref{fig:one}a. \\
We connect these tipping elements via a linear coupling term such that Equation~\ref{fig:one} becomes
\begin{equation}
    \frac{dx_i}{dt}=-a(x_i-x_0)^3+b(x_i-x_0)+c_i + r\sum_{j=1,j\neq i}^N A_{ij}x_j,
\label{eq:network}
\end{equation}
where $r>0$ is the global \textit{coupling strength} between the elements and $A_{ij}$ is one if there exists a link from node $j$ to $i$ and it is zero otherwise. Thus, the networks considered here are directed, however, couplings of the node to itself are not considered. In our network, we use $a=4$, $b=1$ and $x_0=0.5$ for all nodes (i.e., tipping elements) such that the stable states $x_i$ are at 0 or 1 respectively if $c_i=0$. If not stated otherwise, we simulate all our results on Erd\H{o}s-Rényi networks~\citep{erdos1959random} of size 100. This means that our work here is based on the network framework developed in Krönke et al. (2019)~\citep{kronke2019dynamics}. However, Krönke et al. (2019)~\citep{kronke2019dynamics} touch on important global features of the model, whereas this work emphasizes how small-scale structures change the behavior of the entire network. Furthermore, the system is assumed to be in equilibrium as soon as the change of the state of no tipping element exceeds $0.005$ per time step.

\subsection{Definition of a tipping cascade}
\begin{figure}[htbp]
\centering
\includegraphics[width=.7\textwidth]{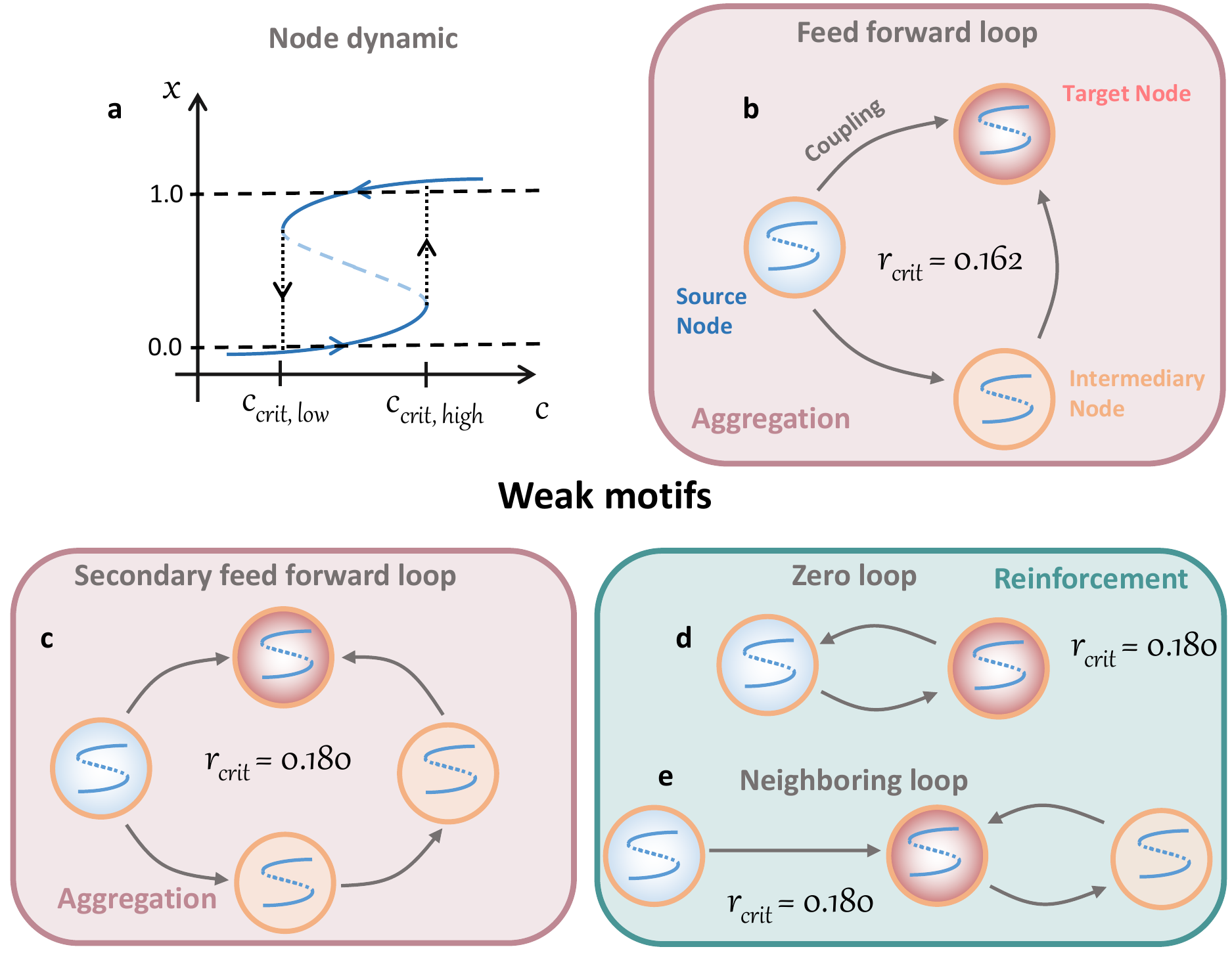}
\caption{\textbf{a)} Bifurcation diagram of a single node (tipping element) in the network. If not stated otherwise, the network size is 100 nodes. Each of these nodes has two stable states, where the stable state is dependent upon the critical parameter $r$. If the critical parameter is increased over a threshold a saddle-node bifurcation occurs. \textbf{b) $-$ e)} Motifs that reduce the minimal necessary critical coupling strength within the network and lead to tipping cascades. In case the source node (light blue) is tipped, the target node (light red) will tip earlier due to the specific local network structure and the additional coupling from the intermediary node(s) and thus triggers a cascade at lower coupling constants than it would be the case if we would only consider a pair of source node and target node. \textbf{b)} Feed forward loop: This motif is a strong motif that reduces the critical coupling strength significantly to 0.162 (from 0.183; see Fig.~\ref{fig:two}). \textbf{c) $-$ e)} Weaker motifs (secondary feed forward loop, zero loop and neighboring loop) that reduce the minimal necessary coupling strength only slightly to around 0.180, where each of the three weaker motifs individually reduces the critical coupling strength to 0.180. The feed forward and the secondary feed forward loop function over \textit{aggregation} effects of coupling, while the zero loop and the neighboring loop function over \textit{reinforcement} feedbacks shown as light red and green colored boxes.}
\label{fig:one}
\end{figure}

In the investigated networks, we define a tipping cascade as the joint transgression of at least two tipping elements in the network. To check if a tipping cascade can occur at a certain coupling strength $r$, a randomly chosen \textit{source node} $i$ is tipped by shifting its individual forcing parameter $c_i$ above its threshold of $c_\text{crit,\ high} = \sqrt{\frac{4}{27}\cdot \frac{b^3}{a}} \approx 0.193$. The individual forcing parameter of all other nodes $c_j$ are kept at zero such that a cascade can only be caused by the coupling of the tipped node to other nodes in the network. With this setting, the cascade simulations in this work are conducted as follows: First, the critical value $c_i$ of source node $i$ is slowly increased (in steps of $0.01$) until $0.193$ is surpassed such that this node tips. Then the simulation is integrated forward in time using python's \textit{scipy.integrate.odeint} until an equilibrium is reached. The equilibrium condition is that $\Delta x_i < 0.005$ in two consecutive time steps for each node $i=1,\, ...,\, N$ in the network. Thus, the cascade simulations are conducted as in Krönke et al. (2019)~\citep{kronke2019dynamics}.\\
Note that if node $i$ tips at $c_\text{crit,\ high,\ i} \approx 0.193$, its stable state in the upper branch is approximately at $x_\text{crit,\ high,\ i} \approx 1.05$, slightly higher than $1.0$ (see Fig.~\ref{fig:one}a). If then node $i$ is coupled to another node $k$ (and no other connections are considered for the moment), the coupling term of Eq.~\ref{eq:network} pointing to node $k$ would be $\text{Cpl}_k = r\sum_{j=1,j\neq k}^N A_{kj}x_j = r \cdot x_\text{crit,\ high,\ i} \approx r \cdot 1.05$. $\text{Cpl}_k$ surpasses the critical value of $0.193$ such that node $k$ would tip as soon as the coupling strength $r$ is larger than $r = r_\text{crit} \approx 0.183$ (see Fig.~\ref{fig:one}a and Eq.~\ref{eq:network}). 

\subsection{Network motifs}
Some of the most important features in networks are small-scale motifs~\citep{Milo2002,gorochowski2018organization}, where a tipped node (\textit{source node}) has a primary direct impact on a \textit{target node}, but also a secondary, indirect impact over \textit{intermediary nodes}. The number of nodes in between a source and a target node over intermediary nodes is called \textit{secondary impact path length}. Thus, a connection of a source node over one intermediary node to a target node would have a secondary impact path length of two. In Fig.~\ref{fig:one} b $-$ e, we show all motifs that have a secondary impact path length of two (\textit{feed forward loop}) and three (\textit{secondary feed forward loop}, \textit{zero loop} and \textit{neighboring loop}). In case of the zero loop, the intermediary node is also the source node. The critical coupling strength of the feed forward loop to tip the target node is reduced from 0.183 to 0.162, for the weaker motifs it is reduced to 0.180 for each of the motifs individually, as we found by simulations. The two types of feed forward loops reduce the critical coupling strength over aggregation effects towards the target node, while the zero loop and the neighboring loop do this via reinforcement loops. The underlying dynamical mechanism is that feed forward loops decrease the critical coupling strength more than weaker motifs. They also contribute more to the average clustering coefficient which is linked to a decrease in critical coupling strength~\citep{kronke2019dynamics}. Hence, on a macro-scale, if there are more feed forward loops, the critical coupling strength decreases, while the clustering increases. This is more of a correlation, not a causation.

\subsection{Real-world application: The Amazon rainforest network}
The Amazon moisture recycling network is a network of atmospheric water flows within the Amazon rainforest. The Amazon can be seen as a network of tipping elements~\citep{kronke2019dynamics,zemp2017self} where forests may locally tip to a state of low tree cover, depending on rainfall levels~\citep{hirota2011global}. If an area contains a forest, evaporation is higher than without a forest, as trees can access deep groundwater which they release to the atmosphere in a process called transpiration. Because this atmospheric water rains down over other parts of the forest, forest transpiration is a mechanism by which tipping elements are connected. This cycling of forest transpiration to rainfall was simulated by Staal et al. (2018)~\citep{staal2018forest} and analyzed as a network~\citep{kronke2019dynamics,zemp2017self}. Our nodes are the forests within areas of a size of 2$\times$2$^\circ$. We use the simulated transpiration flows between these nodes for 2014. For further details on the methods behind the simulations, we refer to Staal et al. (2018)~\citep{staal2018forest}.

\section{Results}
\subsection{Motifs in sparse networks}
\begin{figure}[htbp]
\centering
\subfigure{\includegraphics[width=.33\textwidth]{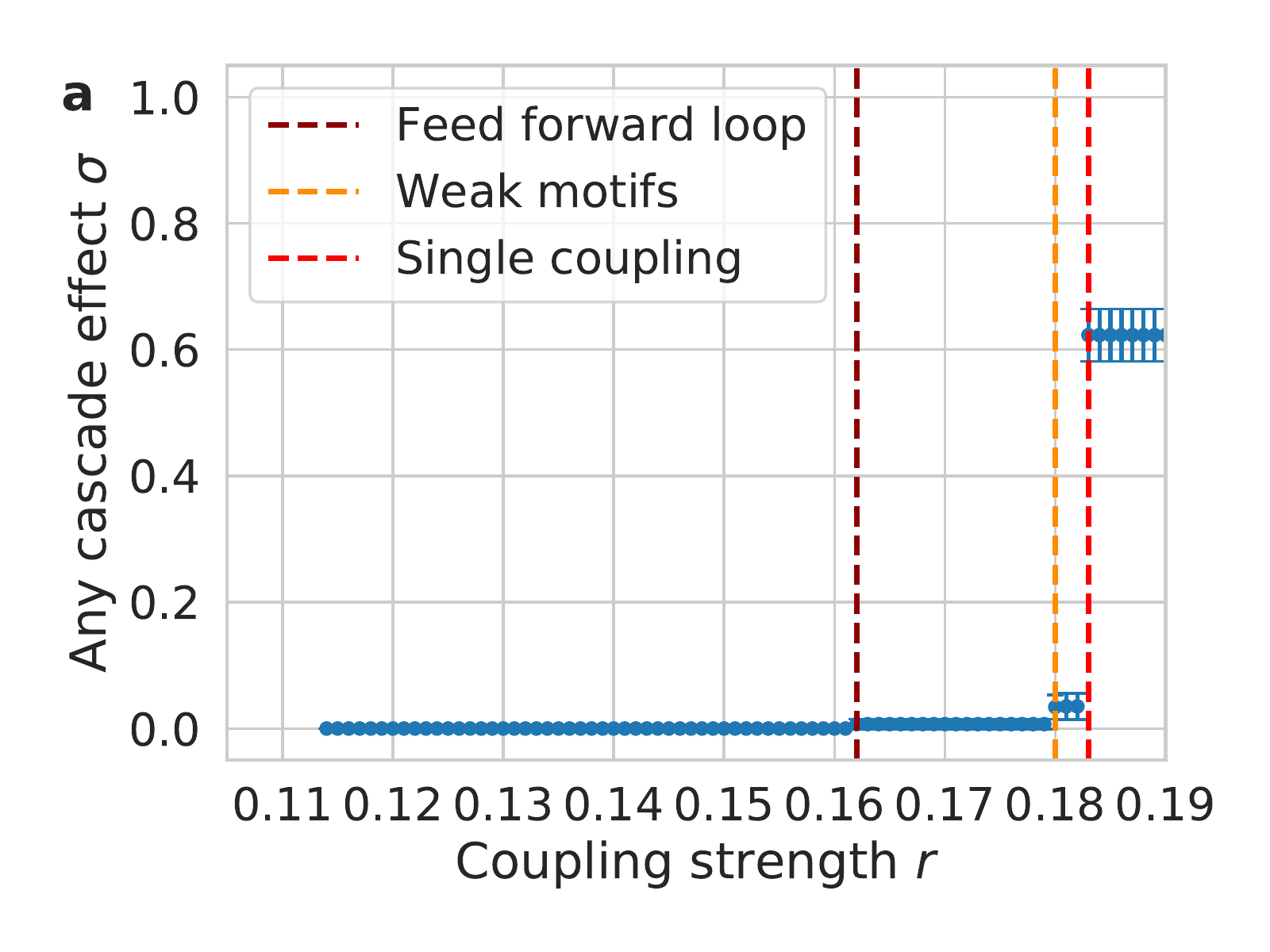}}
\subfigure{\includegraphics[width=.33\textwidth]{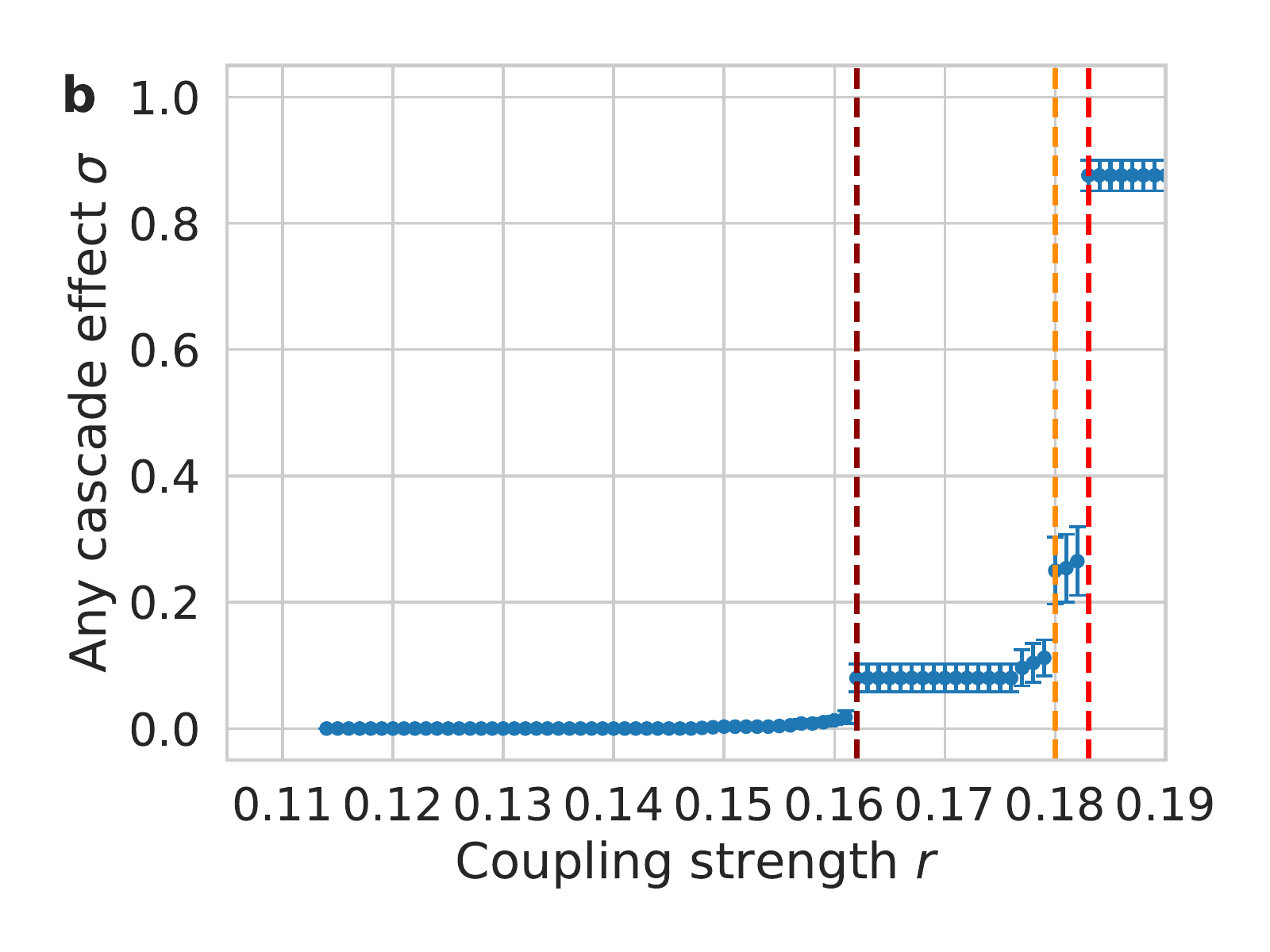}}\\
\subfigure{\includegraphics[width=.33\textwidth]{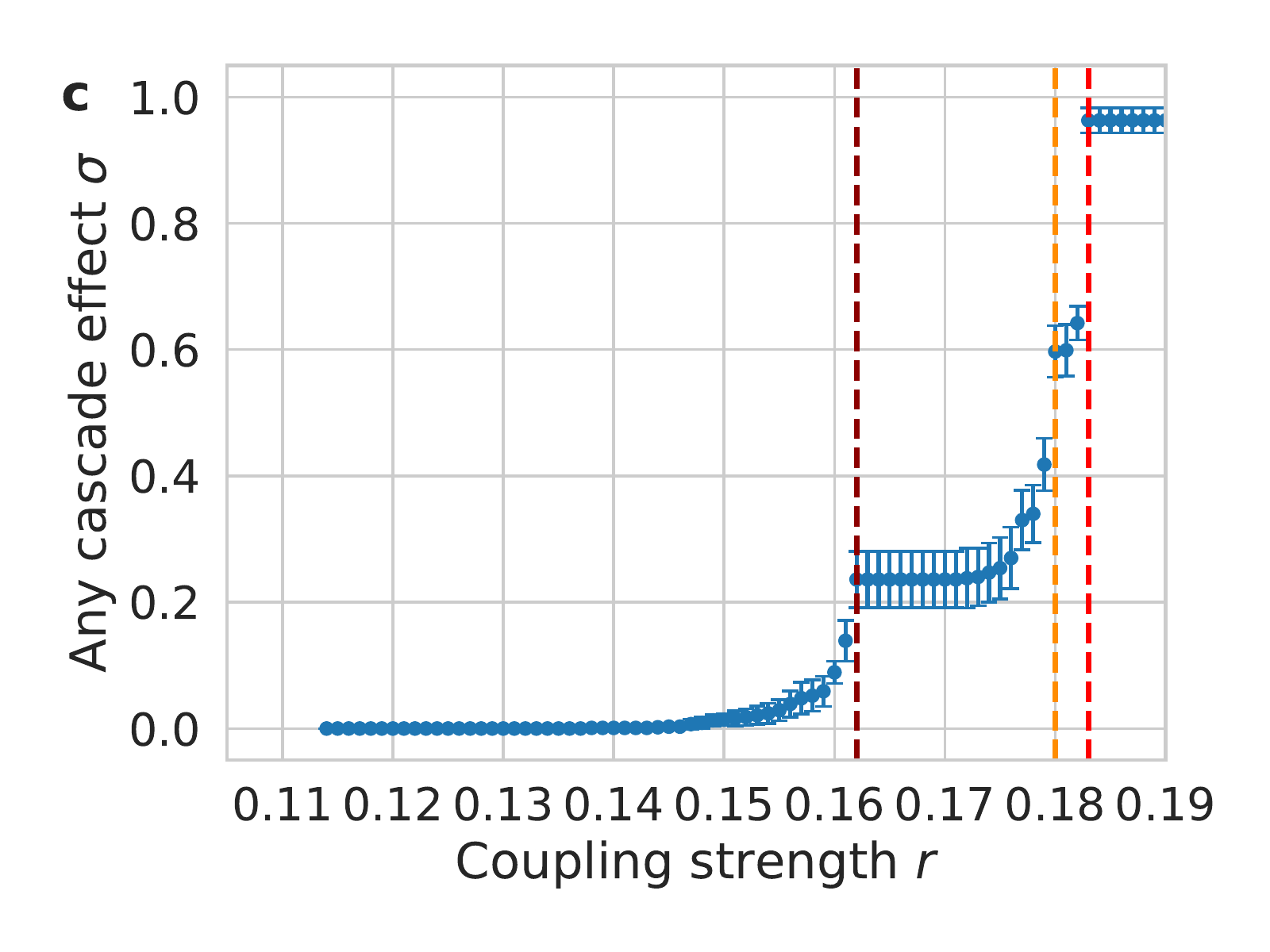}}
\subfigure{\includegraphics[width=.33\textwidth]{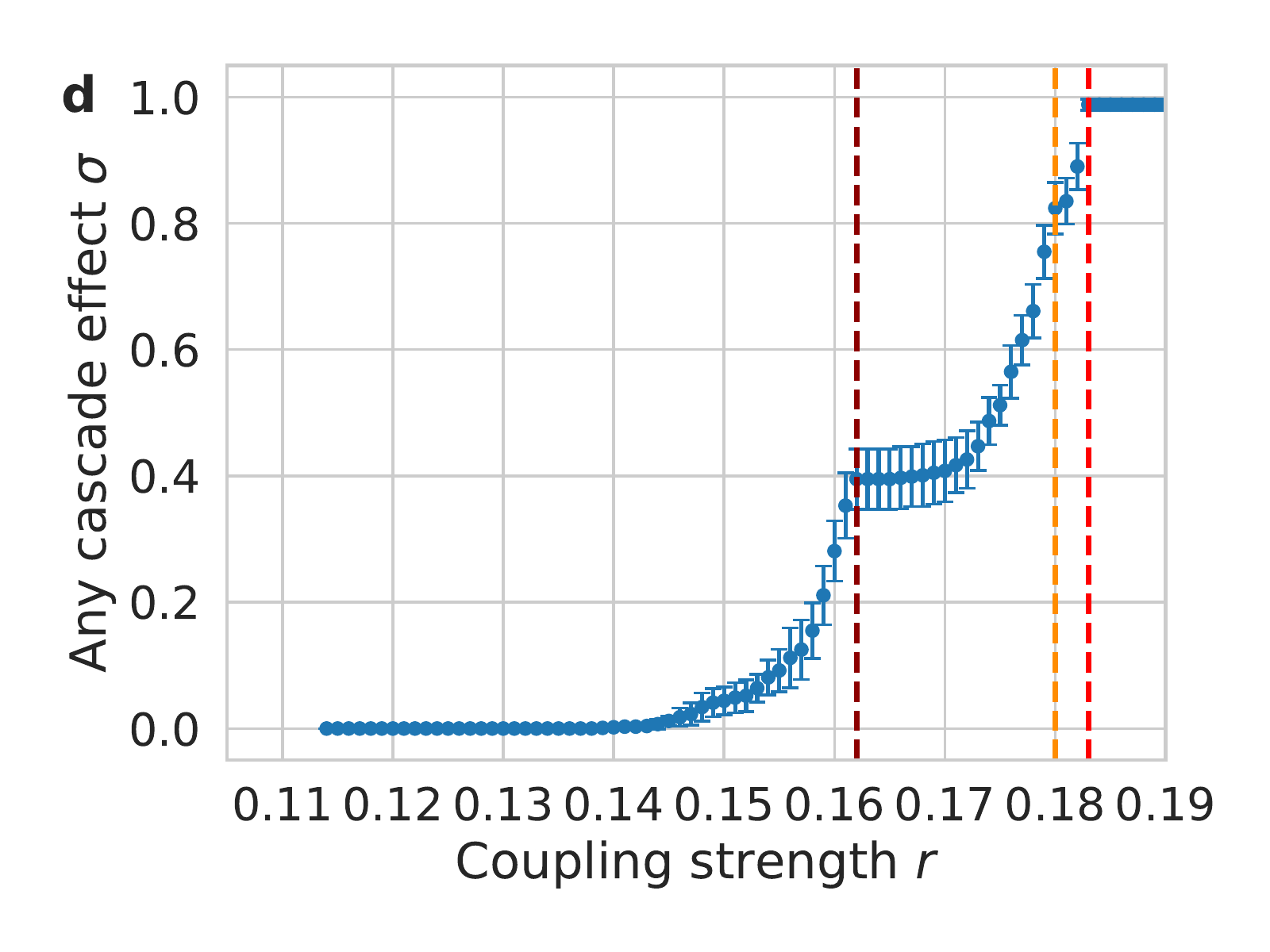}}\\
\subfigure{\includegraphics[width=.33\textwidth]{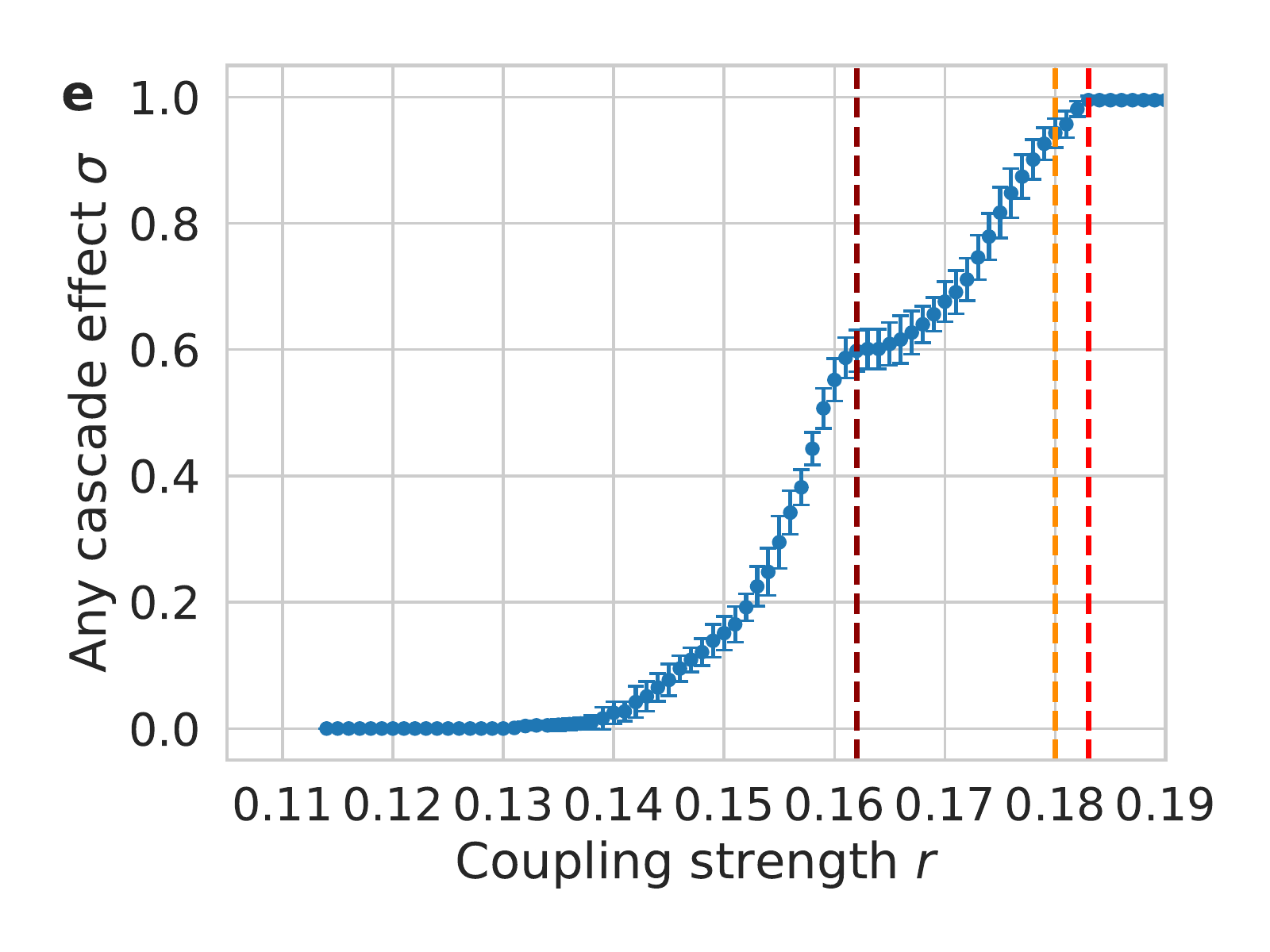}}
\subfigure{\includegraphics[width=.33\textwidth]{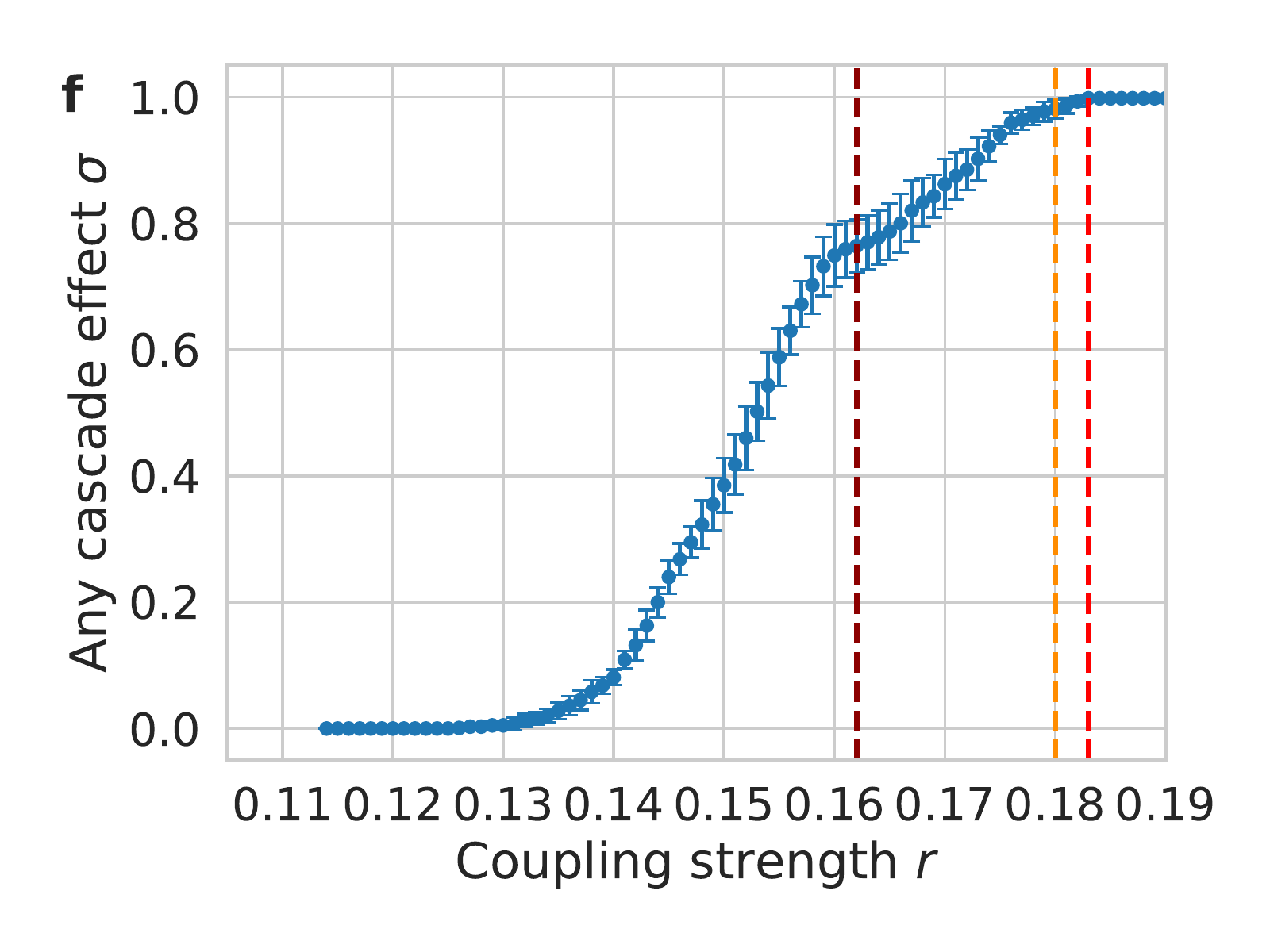}}\\
\subfigure{\includegraphics[width=.33\textwidth]{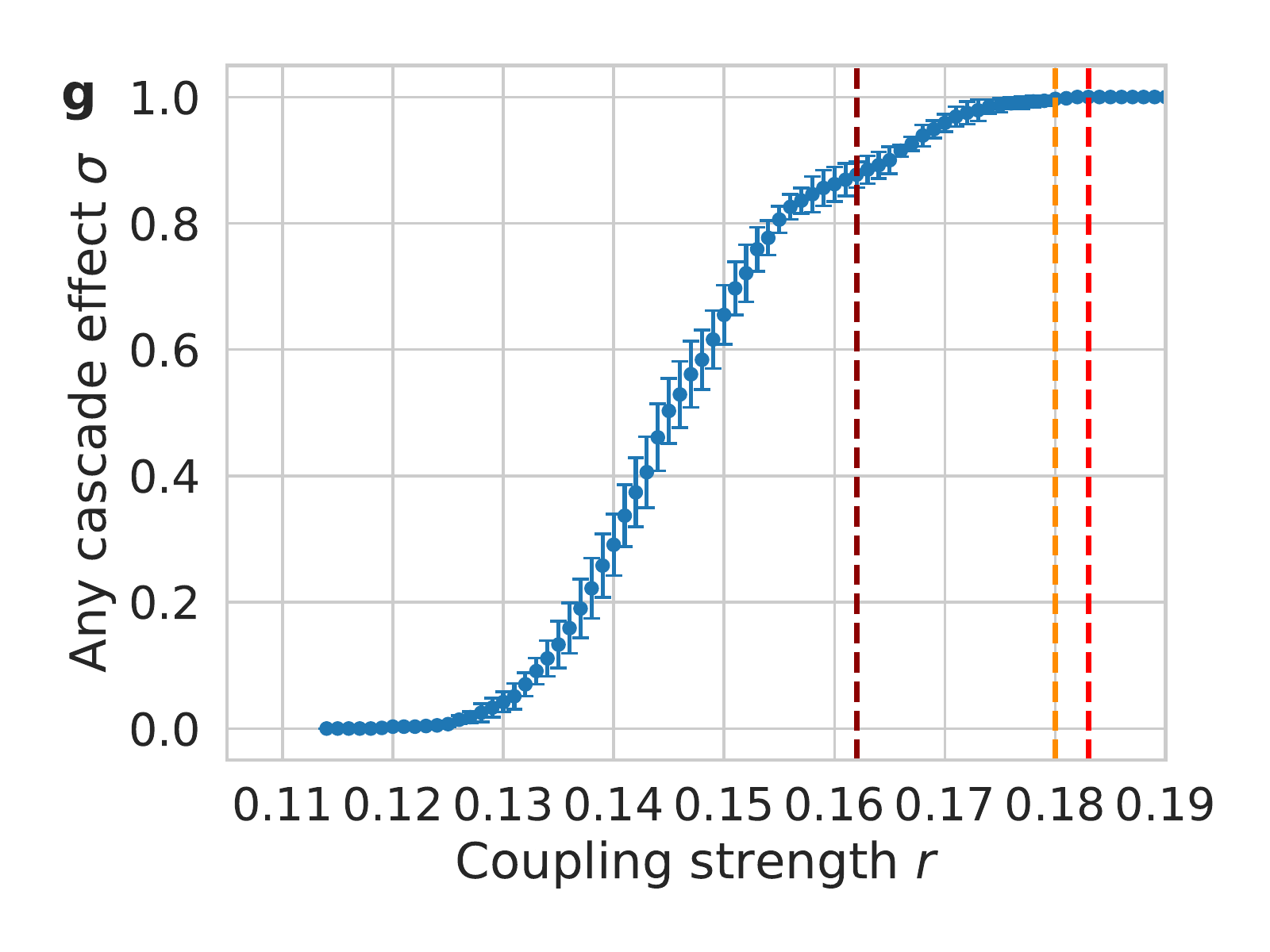}}
\subfigure{\includegraphics[width=.33\textwidth]{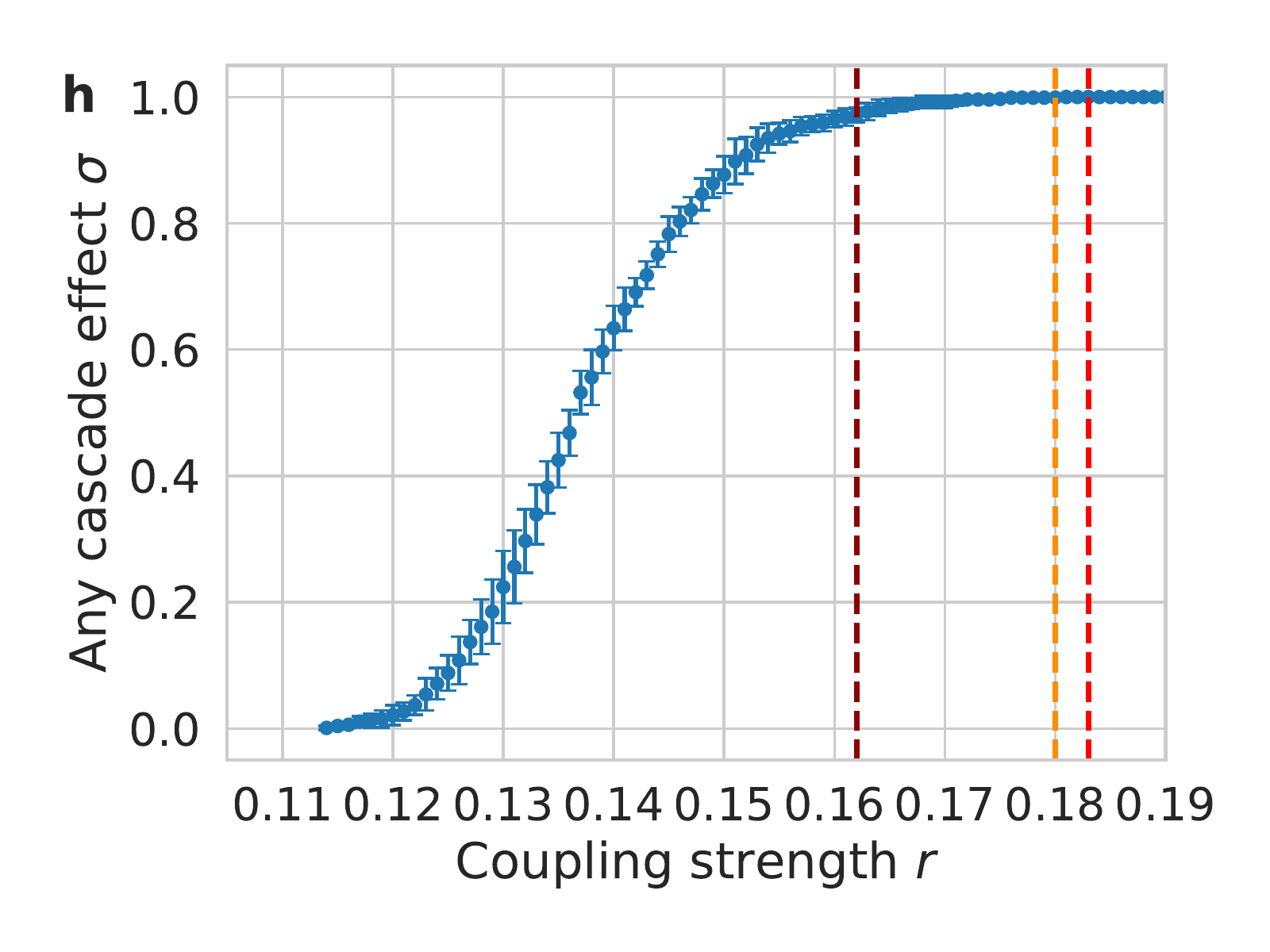}}
\caption{The effect of the coupling strength on the proportion of networks that show any cascading effect. The critical coupling strength to start a cascade for the feed forward loop is 0.162, for the three weaker motifs 0.180 and for the single coupling 0.183. Since each of the weaker motifs reduces the critical coupling strength to 0.180, it is not possible to separate these three motifs from its tipping cascade pattern. The error bars show the standard deviation of 10 simulations à 100 simulated networks. In total 1000 runs with Erd\H{o}s-Rényi networks of size 100 were computed. \textbf{a)} has average degree 1, \textbf{b)} has average degree 2, \textbf{c)} has average degree 3, \textbf{d)} has average degree 4, \textbf{e)} has average degree 5, \textbf{f)} has average degree 6, \textbf{g)} has average degree 7, \textbf{h)} has average degree 8.}
\label{fig:two}
\end{figure}

We find that particularly in Erd\H{o}s-Rényi networks, motifs can significantly reduce the critical coupling strength that is necessary to start a cascade. In Fig.~\ref{fig:two}, the occurrence of cascades $\sigma$ is shown versus the coupling strength $r$, where vertical lines indicate the coupling strength where a tipping cascade is expected for the respective motif or motif group. The actual fingerprint of the respective motif can be observed in step-like features in cascade occurrences $\sigma$ towards higher coupling strengths. If the network has an average degree of one, two or three, these reductions can be seen clearly for the feed forward loop as well as for the weak motifs (secondary feed forward loop, zero loop and neighboring loop; see Fig.~\ref{fig:two}a, b, c). Towards higher average degrees two things can be found: first, cascade occurrence increases and second, the coupling strength $r$ at which cascade occurrences are different from zero decreases. For instance at an average degree of eight, cascades can already be found for a coupling strength around 0.12, whereas for an average degree of two, this coupling strength is around 0.16 (see Fig.~\ref{fig:two}b and h). This might be due to the fact that combinations of different or the same motifs point to the same target node (see for instance supp. Fig. S3). Since the reduction of the critical coupling strength for the feed forward loop is larger than for the weaker motifs, it remains visible up to higher average degrees ($<k>$=7; see Fig.~\ref{fig:two}). With increasing average degree, the networks show an increasing likelihood of cascade occurrences and size of cascade (supp. Fig. S1).
The frequency of cascades does not reach 100\% for average degree one (around 60\%), two (about 90\%) and three (about 99\%) even if the coupling is above 0.183, the coupling value at which a pair of two tipping elements tip (\textit{Single coupling} in Fig.~\ref{fig:two}). The reason is that for low average degrees, the Erd\H{o}s-Rény network is not in the connected regime, meaning not all nodes are part of the giant component. Consequently for low average degrees, some nodes cannot be involved in the tipping cascade as they do not hold any couplings, i.e., their in-degree is zero.\\
\begin{figure}[htbp]
\centering
\subfigure{\includegraphics[width=.49\textwidth]{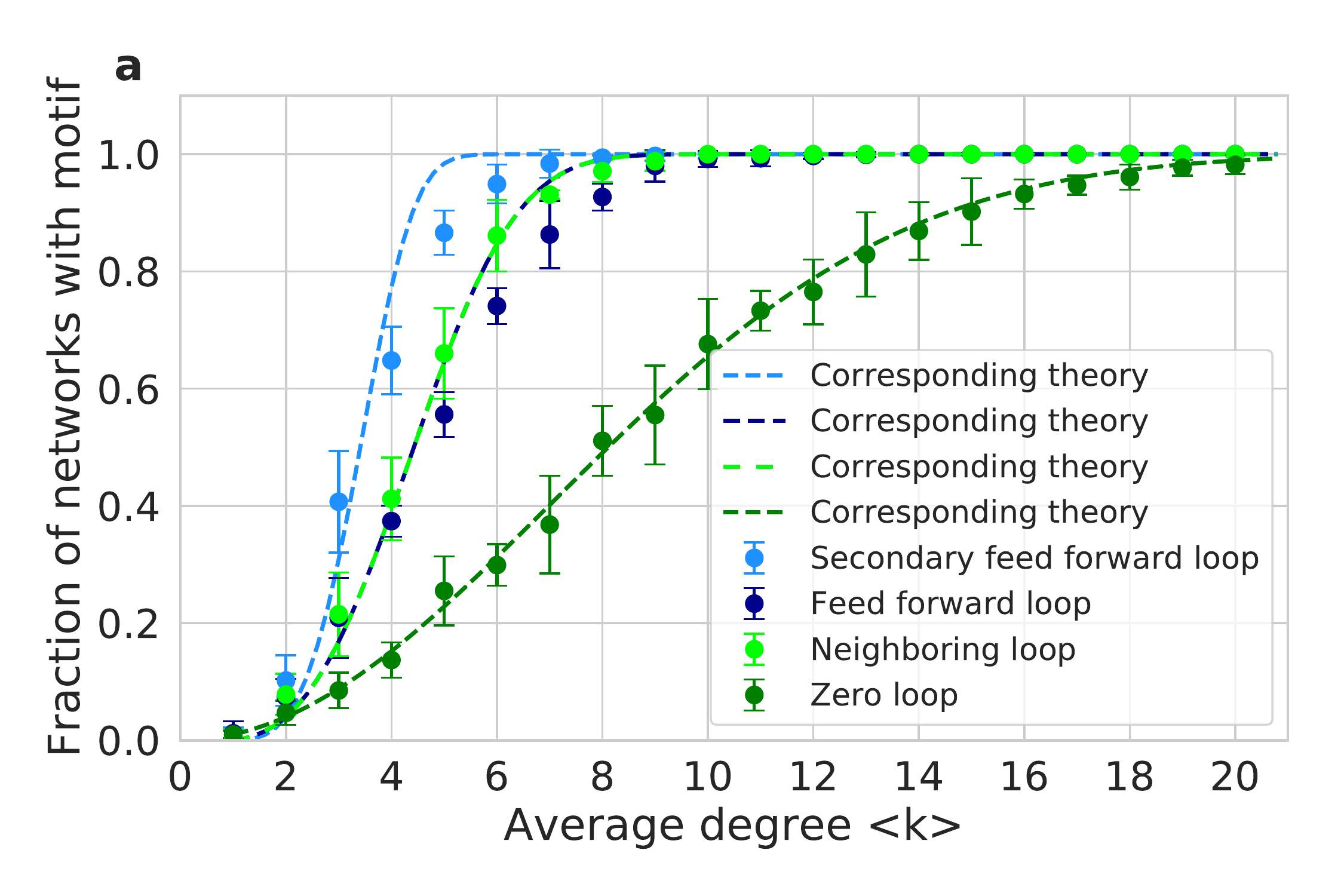}}
\subfigure{\includegraphics[width=.49\textwidth]{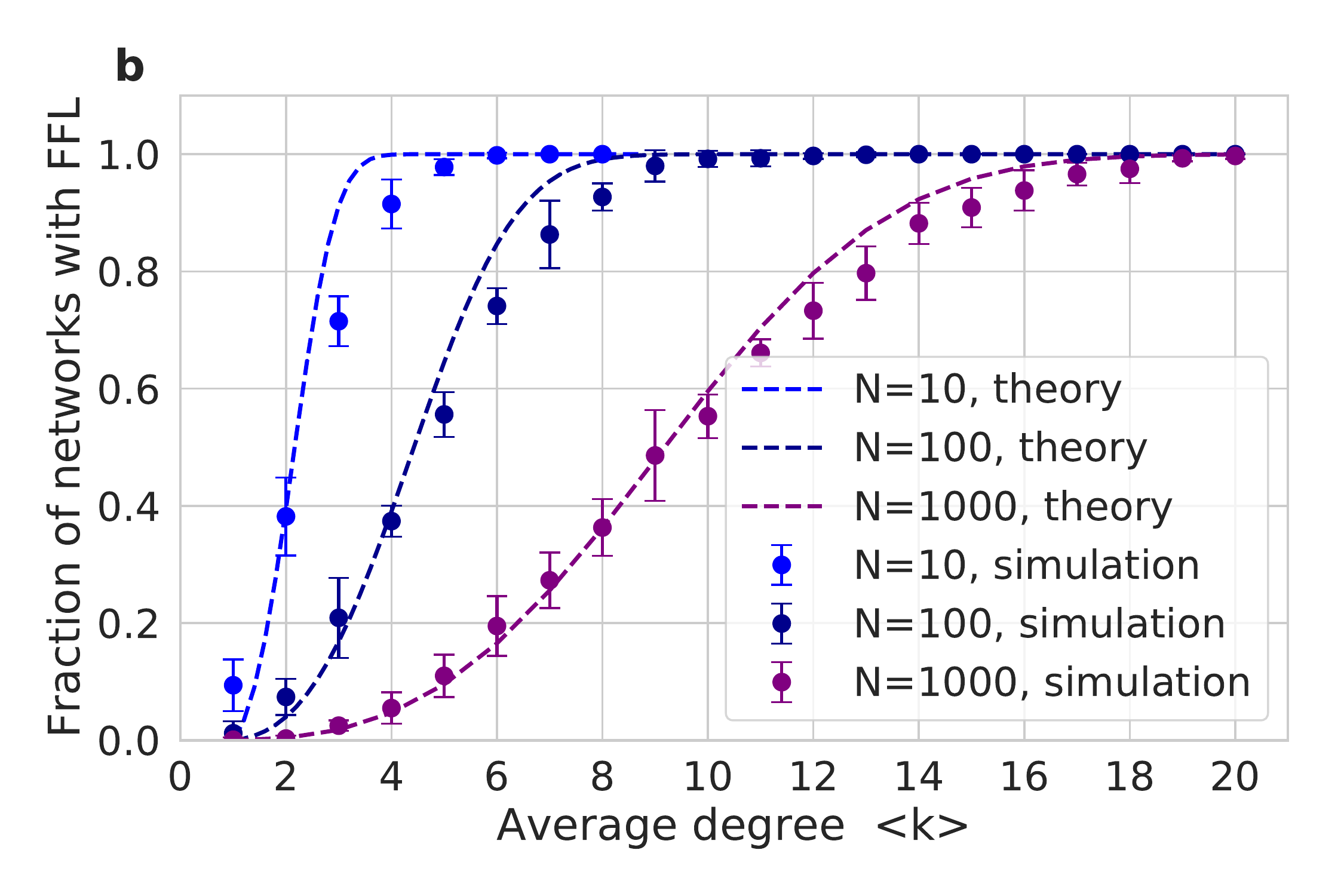}}
\caption{\textbf{a)} Motif occurrence in a random network of size 100 together with their theoretically expected values (dashed lines). The theoretical formulae are found in Equations~\ref{eq:FFL},~\ref{eq:NL},~\ref{eq:ZL}, and~\ref{eq:SFFL}. \textbf{b)} Scaling of the occurrence of the feed forward loop for networks of size 10, 100 and 1000. Theory from Eq.~\ref{eq:FFL}. The other motifs' scaling behavior can be found in supp. Fig.~S2. Error bars in both panels show the error in occurrence in 1000 realisations grouped as 10 $\times$ 100 samples.}
\label{fig:three}
\end{figure}
How often are motifs expected in random networks? The proportion of networks with the respective motif depending on the average degree is sharply ascending (Fig.~\ref{fig:three}a). Here, we compare the simulation (points and error bars) to the theory (dashed lines) and obtain a good match. In the simulation, the occurrence is the probability to find the respective motif at an arbitrarily chosen node.\\
The theoretically derived values can be obtained with the following considerations: for an Erd\H{o}s-Rényi network with average degree $\left<k\right>$, each node is expected to have $\left<k\right>$ neighbors that it is linked to. Thus, the number of possible pairs between any two neighbor nodes is given by ${\left<k\right> \choose 2} = \frac{\left<k\right> \cdot \left( \left<k\right> -1 \right) }{2}$. In a directed network, this number needs to be multiplied by $2$ such that the number of possible links is given by $N_{\text{possible\ links}} =\left<k\right> \left( \left<k\right> - 1 \right)$. The probability that at least one event $A_i$ occurs out of $\mathcal{N}$ independent events is given by:
\begin{equation}
    \mathcal{P}\left( \bigcup\limits_{i=1}^\mathcal{N} A_i \right) = 1 - \left( 1- p \right)^\mathcal{N}
\end{equation}
for a fixed probability $p$ that one independent event occurs. In an Erd\H{o}s-Rényi network $p = \frac{\left<k\right>}{N-1}$, where $N$ is the size of the network. This leads to 
\begin{eqnarray}
    \mathcal{P}_\text{feed\ forward\ loop} = 1 - \left( 1- p \right)^{N_\text{possible\ links}} = 1 - \left( 1- \frac{\left<k\right>}{N - 1} \right)^{\left<k\right> \left( \left<k\right> - 1 \right)} 
    \label{eq:FFL}
\end{eqnarray}
to have at least one feed forward loop at any node of the network. Similarly this approach can be used for the neighboring loop. There are on average $ \left< k \right> \left( \left< k \right> - 1 \right)$ possibilities that neighbor to neighbor nodes form a feedback such that it results in a neighboring loop. Accordingly we have
\begin{eqnarray}
    \mathcal{P}_\text{neighboring\ loop} =  \mathcal{P}_\text{feed\ forward\ loop} = 1 - \left( 1- \frac{\left<k\right>}{N - 1} \right)^{\left<k\right> \left( \left<k\right> - 1 \right)}. 
    \label{eq:NL}
\end{eqnarray}
For the zero loop $\left<k\right>$ possible links that need to be considered for reconnecting any neighbor node back to the source node such that the probability of at least one occurrence is given by
\begin{equation}
    \mathcal{P}_\text{zero\ loop} =  1 - \left( 1- \frac{\left<k\right>}{N - 1} \right)^{\left<k\right>}.
    \label{eq:ZL}
\end{equation}
Finally, we compute the probability for the secondary feed forward loop. We know that the number of neighbors-of-neighbors is $\left<k\right> \left( \left<k\right> - 1 \right)$ excluding the source node as a neighbor. Each of these neighbors-of-neighbors has $\left( \left<k\right> - 1 \right)$ possibilities to link to a specific target node such that we get
\begin{equation}
    \mathcal{P}_\text{secondary\ feed\ forward\ loop} =  1 - \left( 1- \frac{\left<k\right>}{N - 1} \right)^{\left<k\right> \left( \left<k\right> - 1 \right)^2}. 
    \label{eq:SFFL}
\end{equation}
This is the probability of obtaining at least one secondary feed forward loop at any given node in the network.\\
The occurrences of the feed forward loop, the neighboring loop and the secondary feed forward loop increase sharper than the occurrence of the zero loop with increasing average degree such that at an average degree of 9 the first three motifs occur in practically every Erd\H{o}s-Rényi network of size 100 (Fig.~\ref{fig:three}a).\\
The simulated occurrences of the motifs match reasonably well with the theory. However, for the feed forward loop and the secondary feed forward loop, our theory slightly overestimates the occurrence of these motifs for an intermediate occurrence probability. This is probably due to the fact that out-degrees smaller or equal to 1 at a certain node are neglected in the respective equations (i.e., in Eqs.~\ref{eq:FFL} and~\ref{eq:SFFL}). But in fact, the source node of both motifs, the feed forward loop and the secondary feed forward loop, requires an out-degree of at least two. Otherwise these motifs cannot exist.\\
The scaling of frequency of the feed forward loop for networks of size 10, 100, 1000 shows that for larger networks, the occurrence of motifs requires higher average degrees, for theory and simulations (Fig.~\ref{fig:three}b and supp. Fig.~S2). The scaling behavior of the other three motifs (zero loop, neighboring loop and secondary feed forward loop) can be found in supplementary Fig. 2. The observed scaling dependency of motif occurrence in Erd\H{o}s-Rényi networks can also be interpreted as a dependency on the clustering coefficient $\mathcal{C}$ of the networks since $\mathcal{C} = \frac{\left< k \right>}{N-1}$ in Erd\H{o}s-Rényi networks, since the clustering coefficient is inversely proportional to the network size $N$.

\subsection{Motifs in dense networks}
\begin{figure}[htbp]
\centering
\includegraphics[width=.65\textwidth]{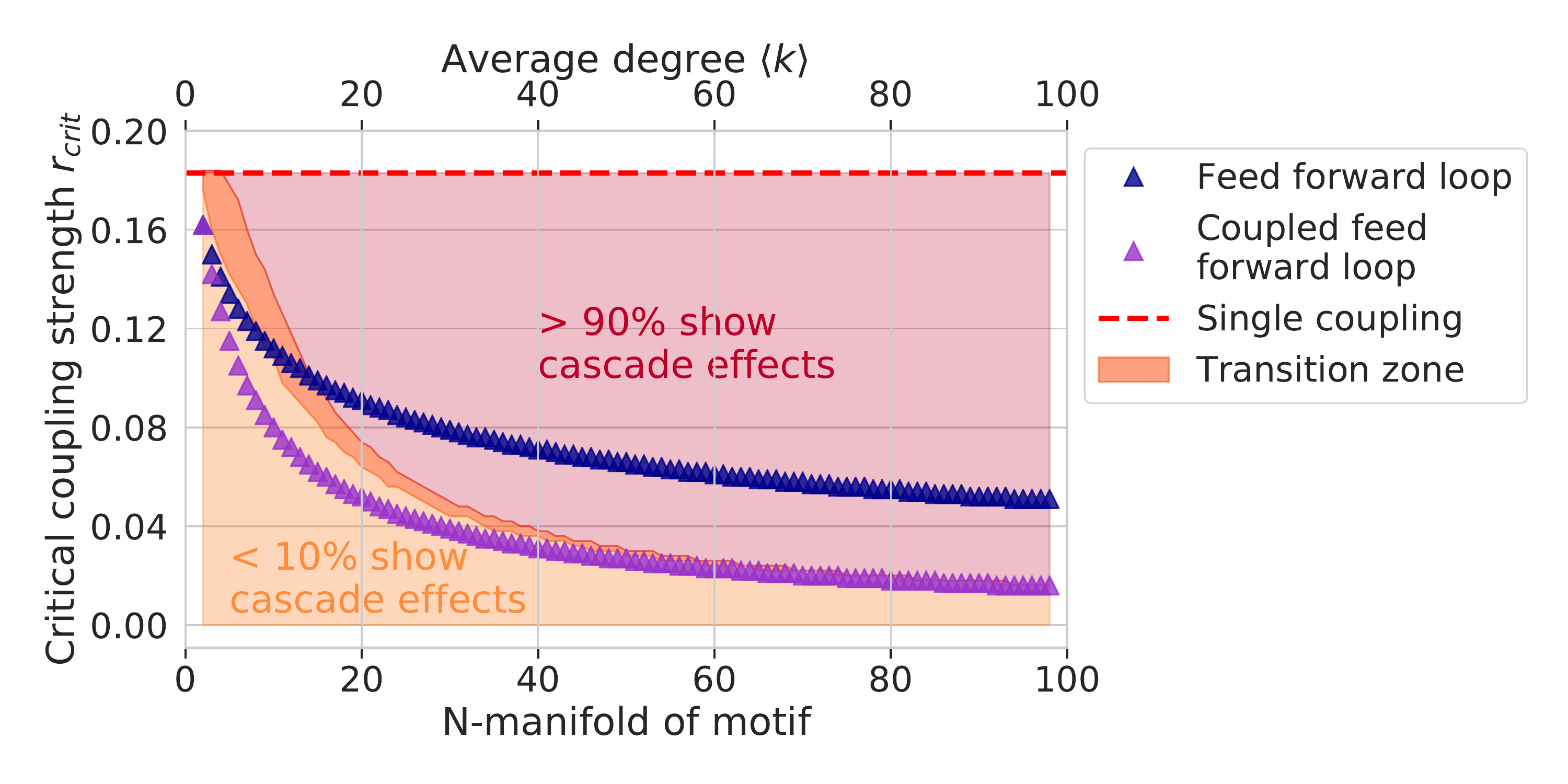}
\caption{The critical coupling strength versus the N-foldness of motifs. The critical coupling strengths are shown as colored triangles for the feed forward loop and the coupled feed forward loop (lower x-axis). The transition zone is the zone where between 10 and 90\% of all networks show a cascading effect (orange shading). Above the transition zone, more than 90\% of all networks show cascading effects (red shading) and below less than 10\% (yellow shading). These measures are shown with respect to the average degree of the network (upper x-axis). For dense Erd\H{o}s-Rényi networks, the match between the transition zone and the N-fold coupled feed forward loop is high. That means that the coupled feed forward loop seems to be a good explanation for the drop of critical coupling strength for densely connected random networks.}
\label{fig:four}
\end{figure}

The occurrence of single motifs plays a crucial role for the occurrence of tipping cascades in sparse networks. For an Erd\H{o}s-Rényi network of size  100, this is the case for average degrees of 6 or below (see Fig.~\ref{fig:two}). However, single motifs cannot explain the drop in critical coupling strength for denser networks. The critical coupling strength for the initiation of cascades lies well below 0.050 in dense networks which is way below the critical coupling strength of a feed forward loop ($r_\text{crit,\ feed\ forward\ loop} = 0.162$); see the transition zone in Fig.~\ref{fig:four}). Above the \textit{transition zone}, more than 90\% of all networks show tipping cascades and below it, less than 10\% show cascades. To explore the strength of the effect of multiple motifs, construction rules for N-fold feed forward loops and N-fold \textit{coupled} feed forward loops were designed (see supp. Fig.~3). Subsequently, numerical simulations of isolated multiple motifs were conducted to assess the critical coupling thresholds (triangles in Fig.~\ref{fig:four}). The isolated, multiple motifs exhibit significantly reduced critical coupling strengths and it can therefore be expected that, in turn, their occurrence in Erd\H{o}s-Rényi networks decrease the critical coupling strength for tipping cascades.\\
The critical coupling strength of a 98-fold coupled feed forward loop ($r_\text{98-fold\ coupled\ feed\ forward\ loop} = 0.016$) matches the critical coupling strength of the transition zone of a fully connected network. This means that the critical coupling compared with the single critical coupling strength of two nodes ($r_\text{crit} = 0.183$) drops by 91\%. It has to be remarked that the critical coupling values of manifold motifs are shown against their multiplicity (lower x-axis in Fig.~\ref{fig:four}), while the critical values corresponding to the transition zone are plotted in relation to the average degree (upper x-axis in Fig.~\ref{fig:four}). Thus, this does not provide direct information which N-fold motif occurs at what average degree, but the comparison between the critical value of the N-fold coupled feed forward loop and the observed critical coupling strength in the Erd\H{o}s-Rényi shows a very good match for networks with high densities of $\left< k \right> > 50$ and is as such a very likely explanation for the observed drop in critical coupling strength.

\subsection{Motifs in a real-world application: The Amazon rainforest}
\begin{figure}[htbp]
\centering
\subfigure{\includegraphics[width=.35\textwidth]{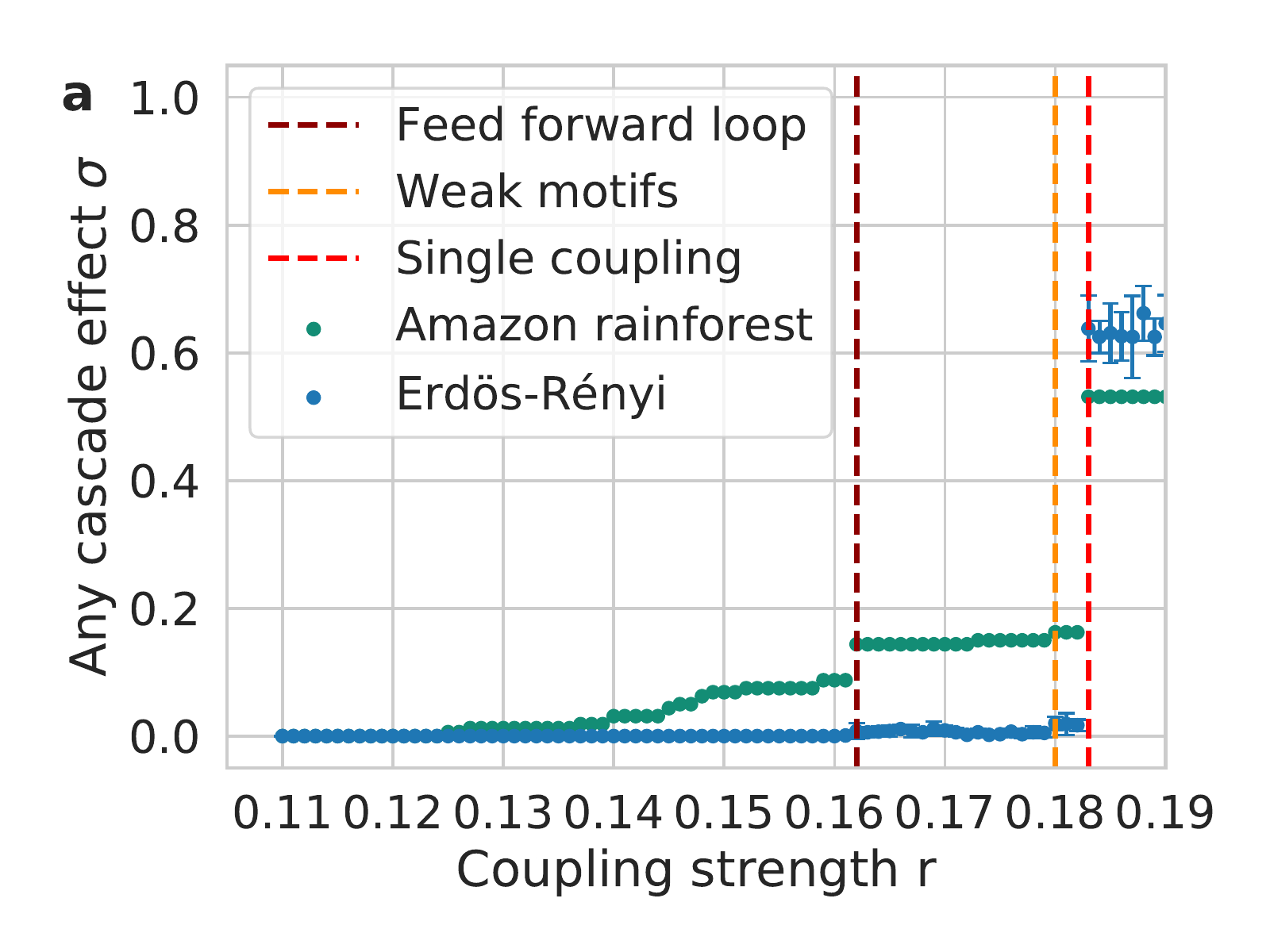}}
\subfigure{\includegraphics[width=.35\textwidth]{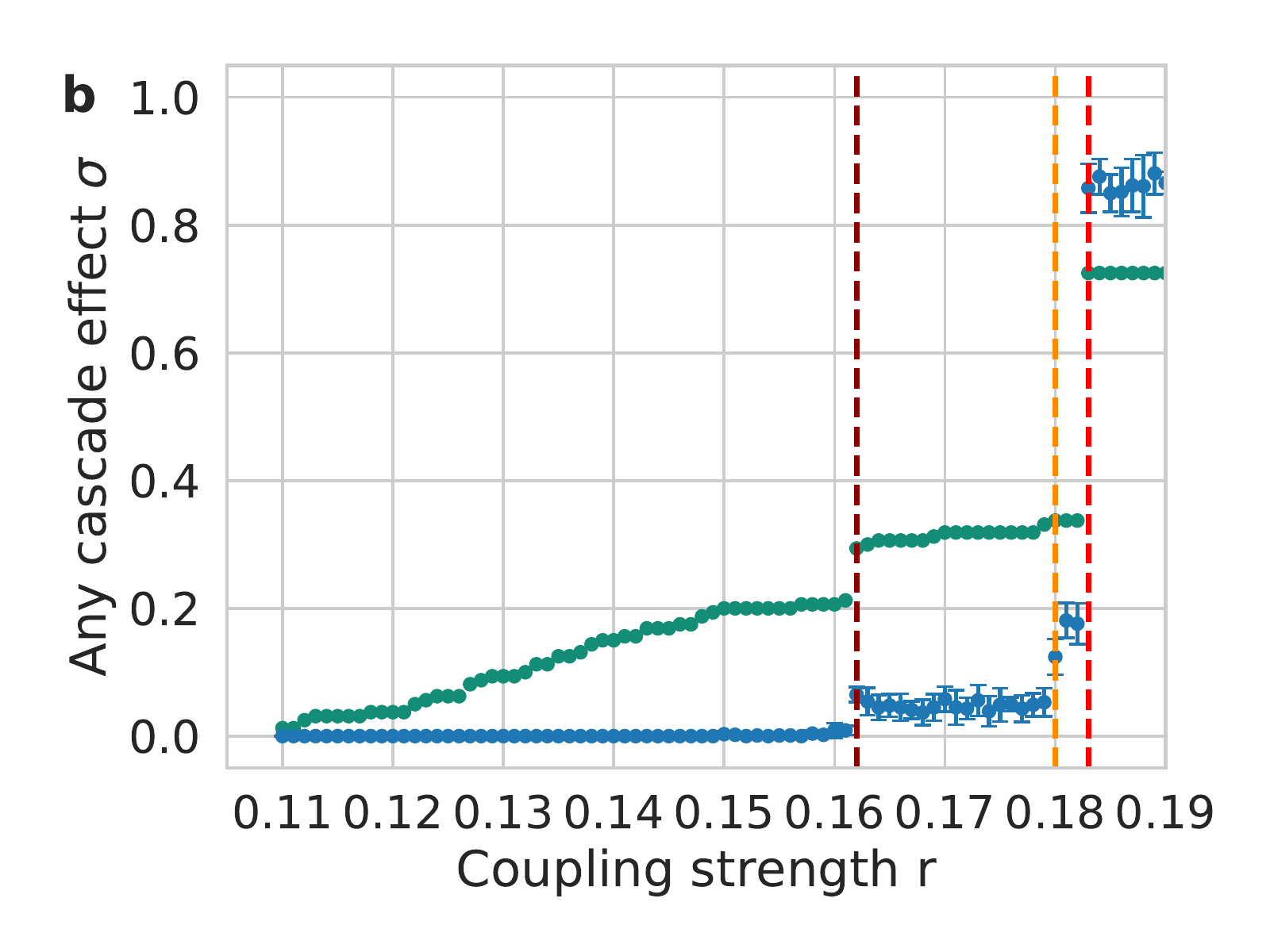}}\\
\subfigure{\includegraphics[width=.35\textwidth]{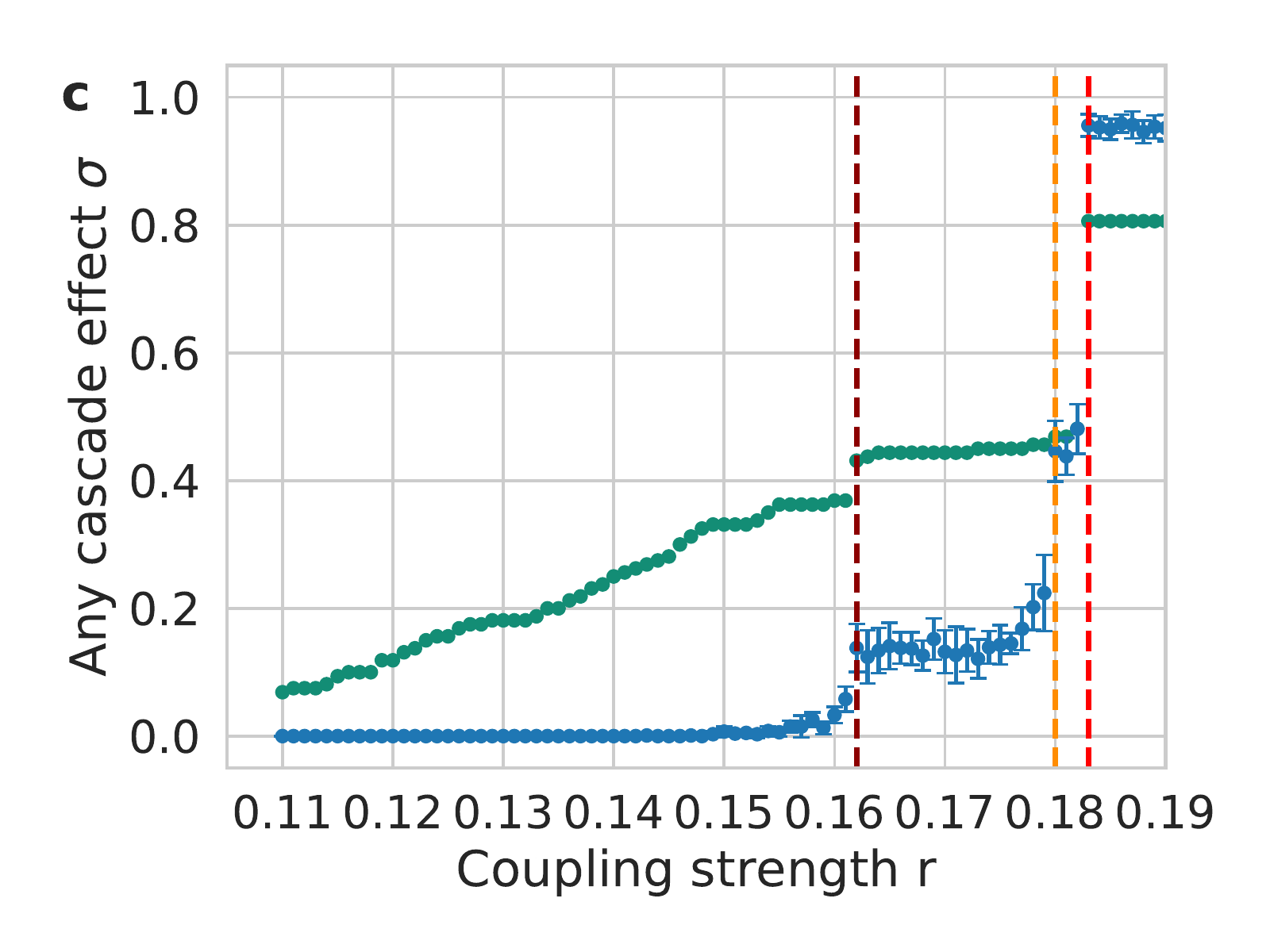}}
\subfigure{\includegraphics[width=.35\textwidth]{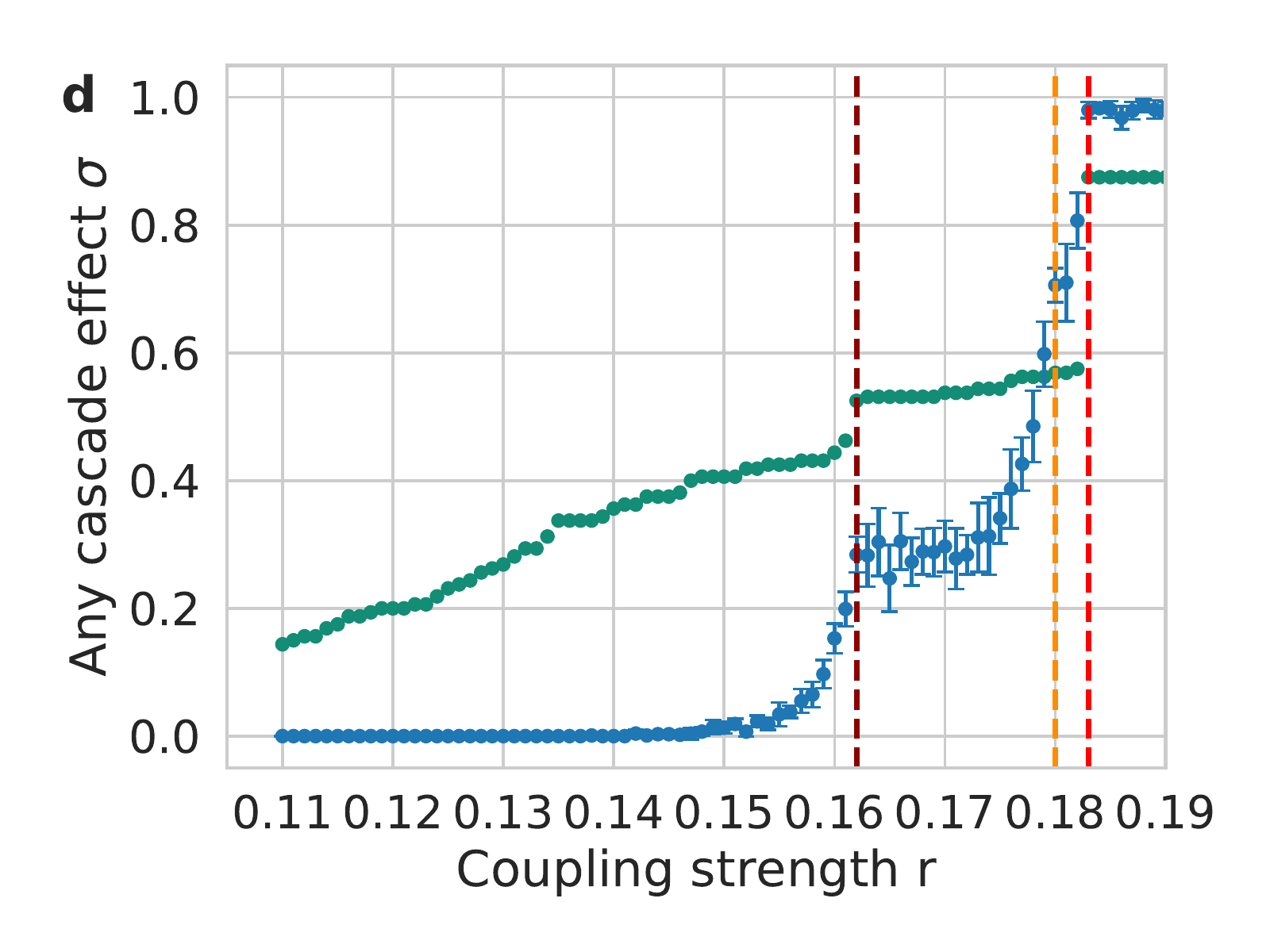}}\\
\subfigure{\includegraphics[width=.35\textwidth]{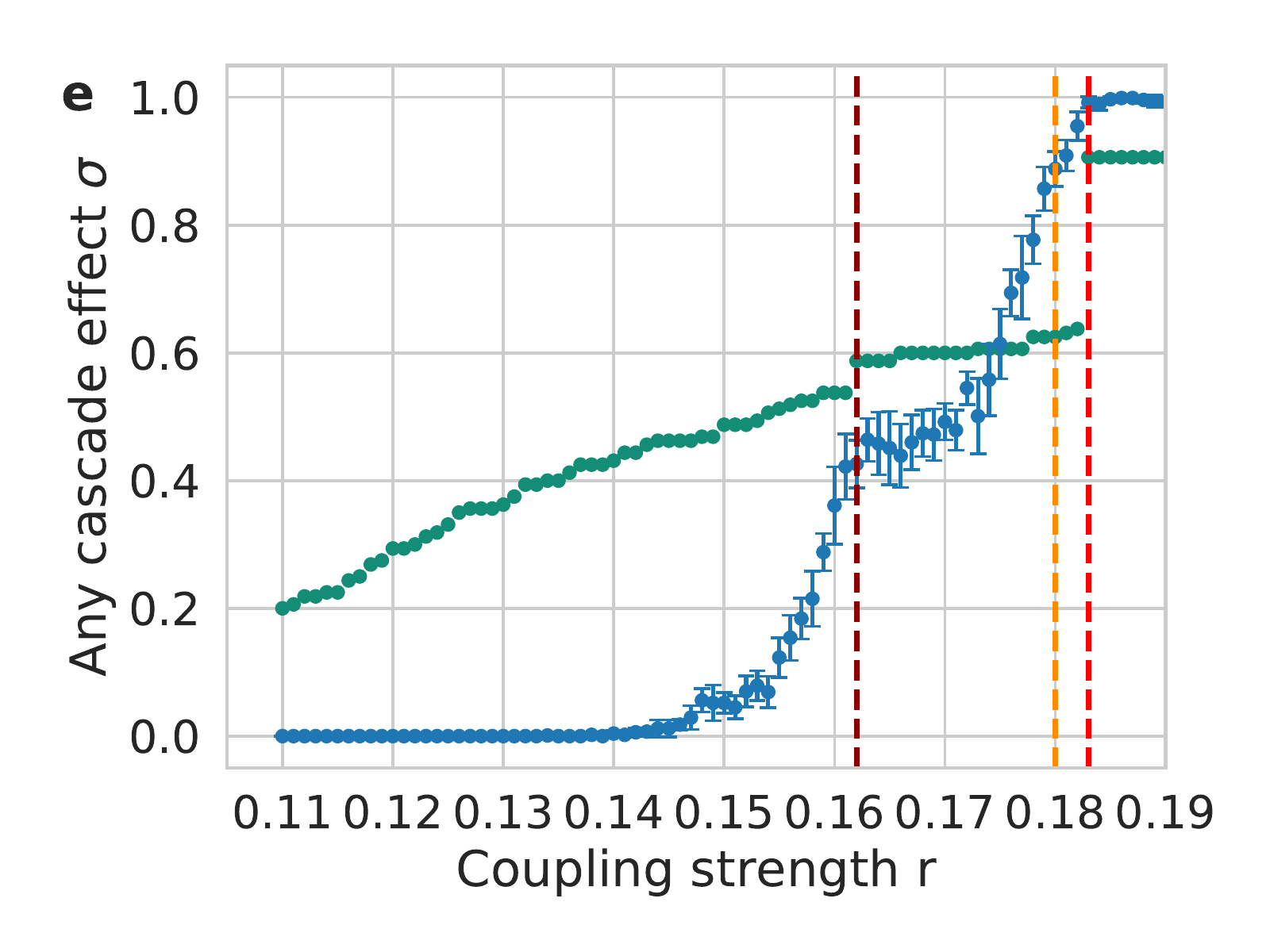}}
\subfigure{\includegraphics[width=.35\textwidth]{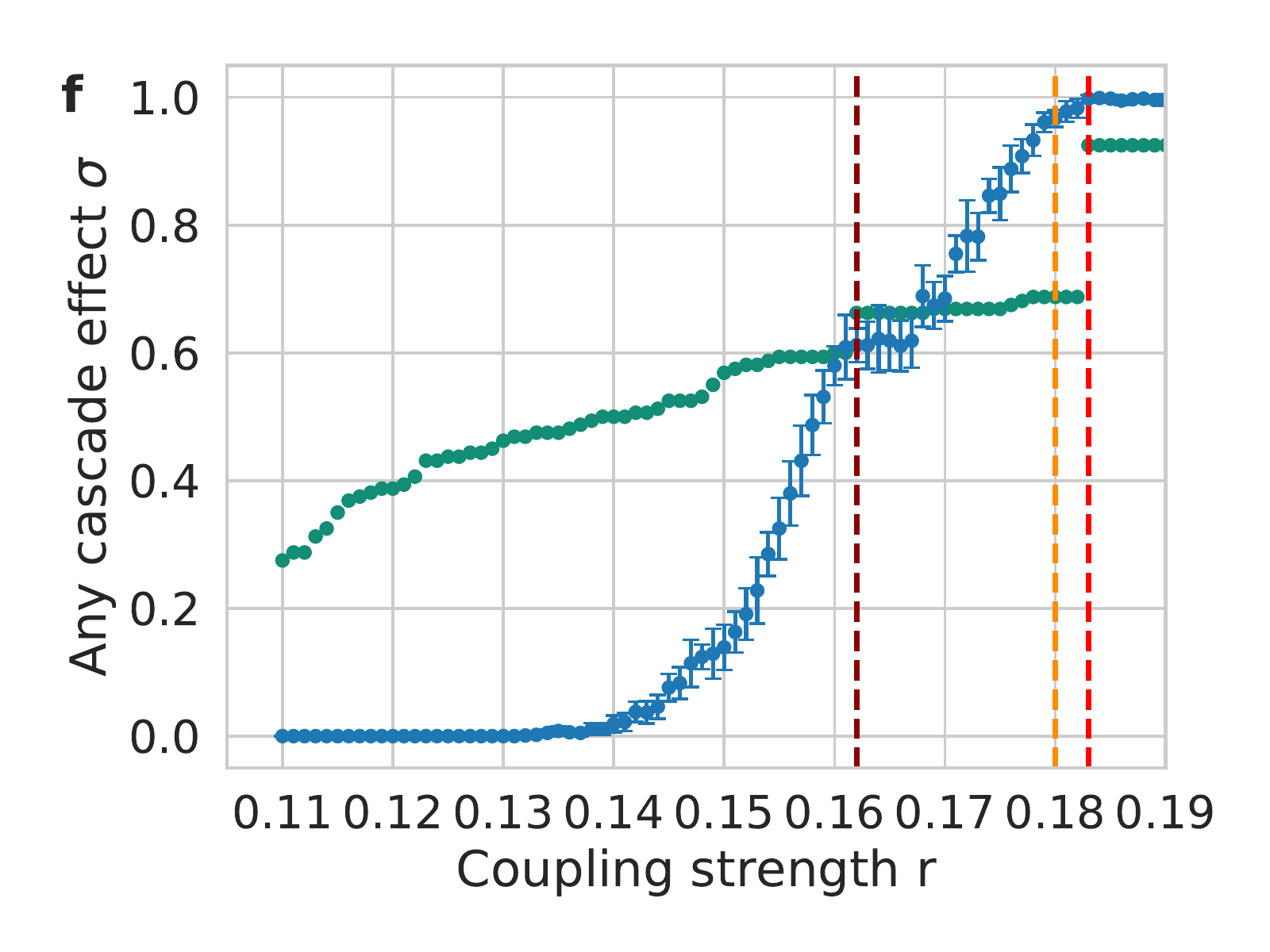}}\\
\subfigure{\includegraphics[width=.35\textwidth]{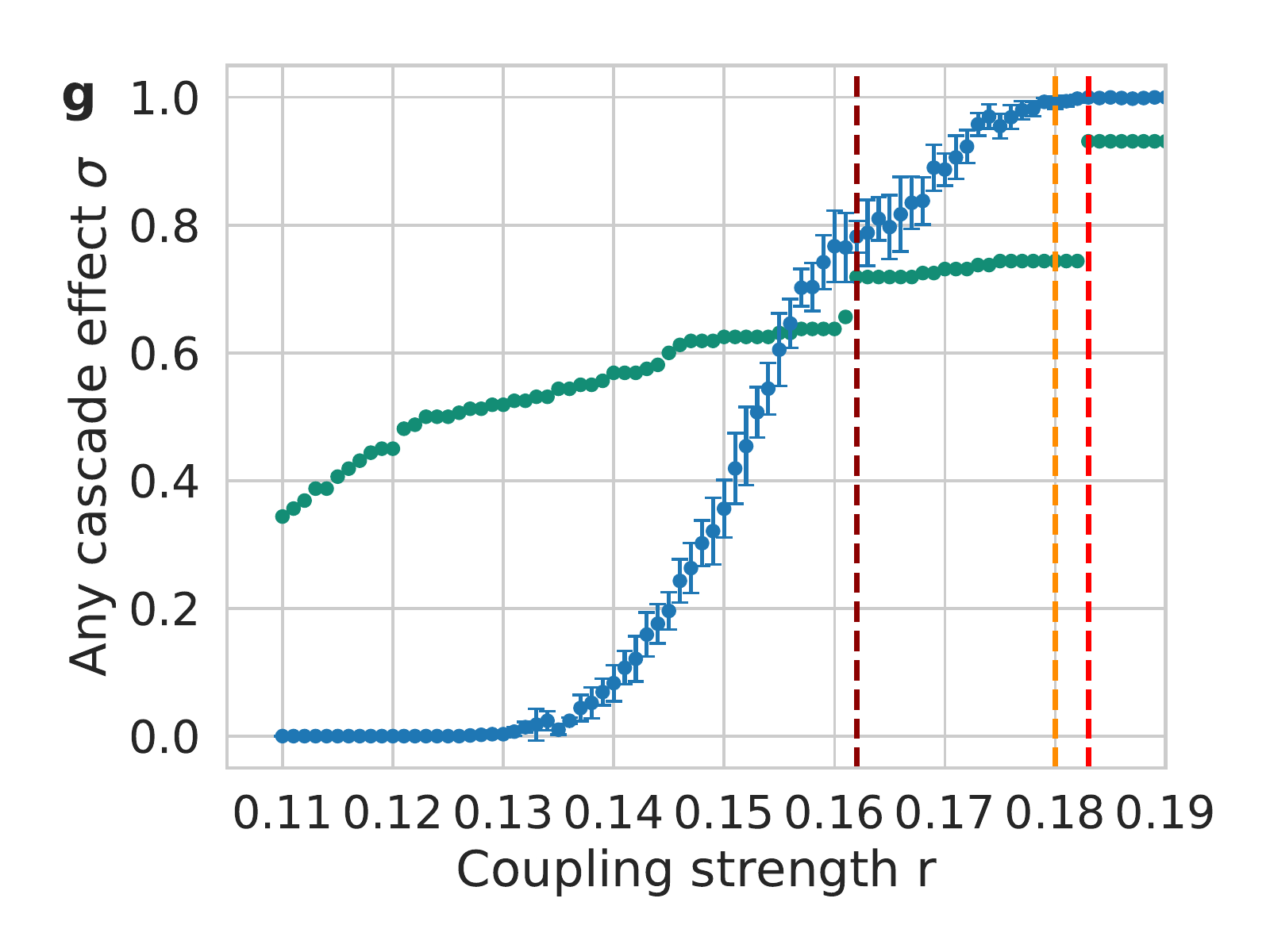}}
\subfigure{\includegraphics[width=.35\textwidth]{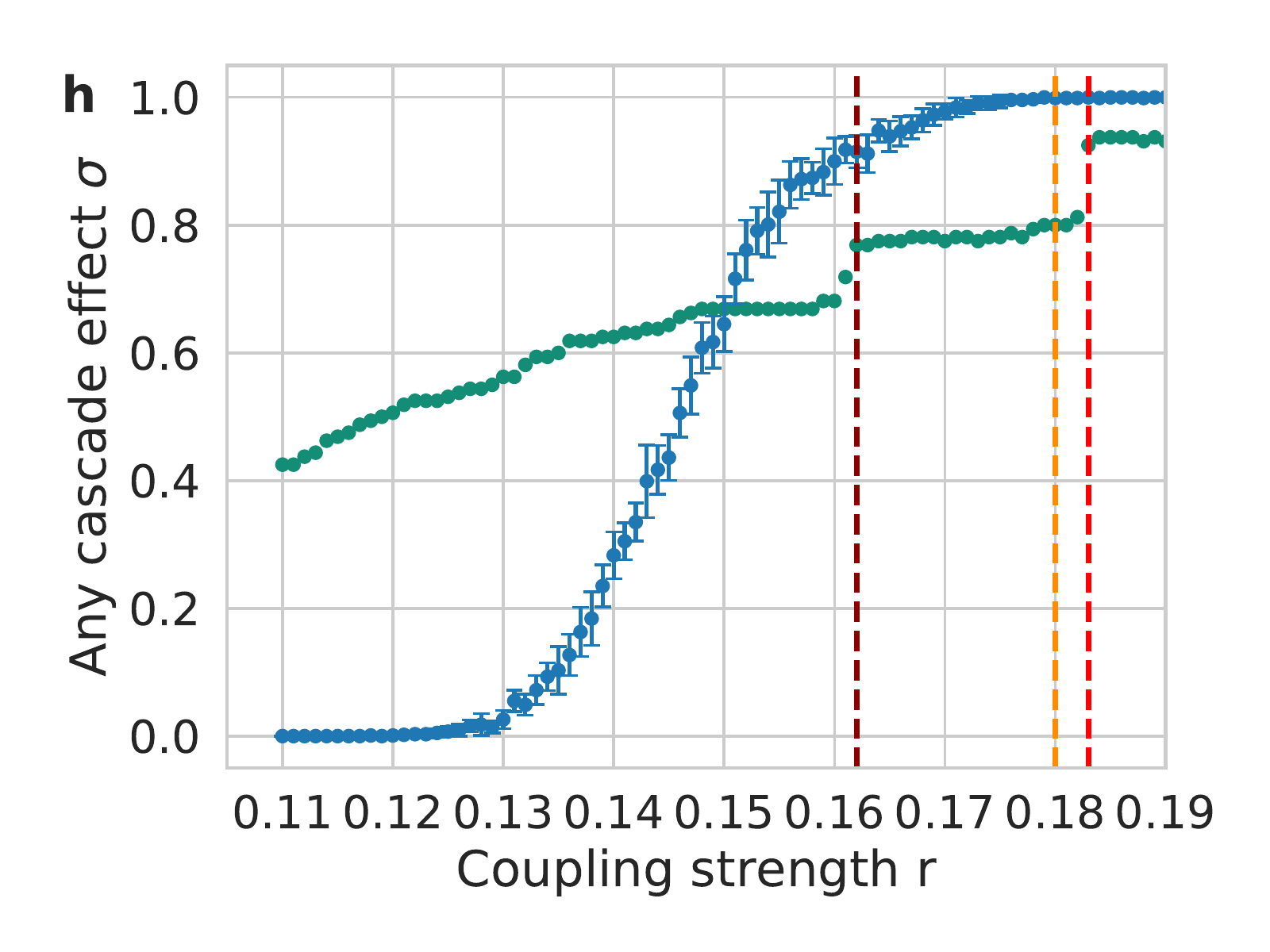}}
\caption{Comparison between tipping cascades in Erd\H{o}s-Rényi networks (blue) and in the Amazon rainforest (green) dependent on the coupling strength $r$ for \textbf{a) $-$ h)} average degrees from 1 $-$ 8. For both network types, the size is 160 nodes. The error for the Erd\H{o}s-Rényi networks is the standard deviation of 10 bundles of simulations, where each bundle consists of 100 tipping cascade experiments (compare to Fig.~\ref{fig:two}). For the moisture recycling network, we simulate cascades for each of the 160 nodes and compute the average number of experiments, where a cascade is observed.}
\label{fig:five}
\end{figure}

\begin{figure}[htbp]
\centering
\subfigure{\includegraphics[width=.45\textwidth]{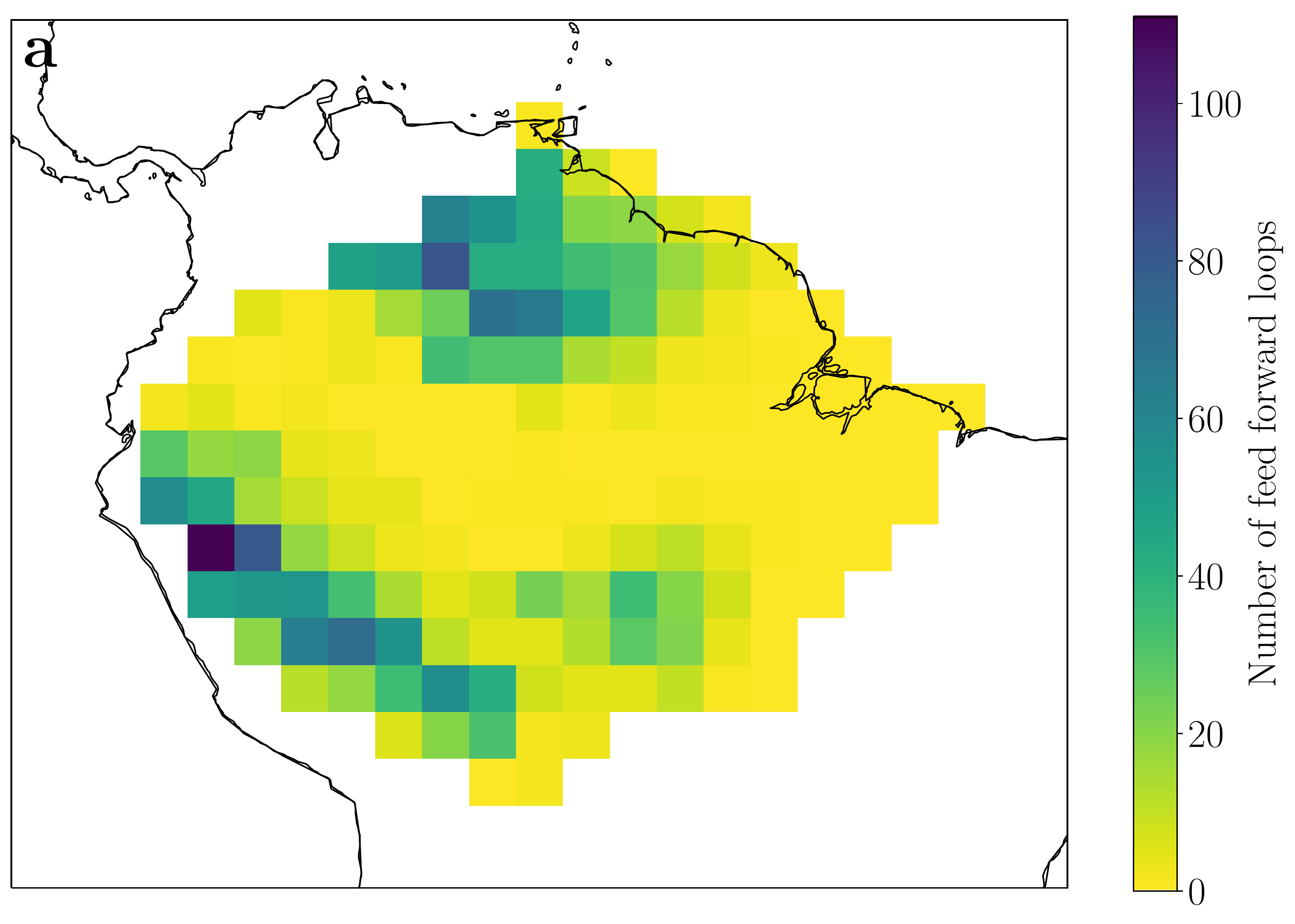}}
\subfigure{\includegraphics[width=.45\textwidth]{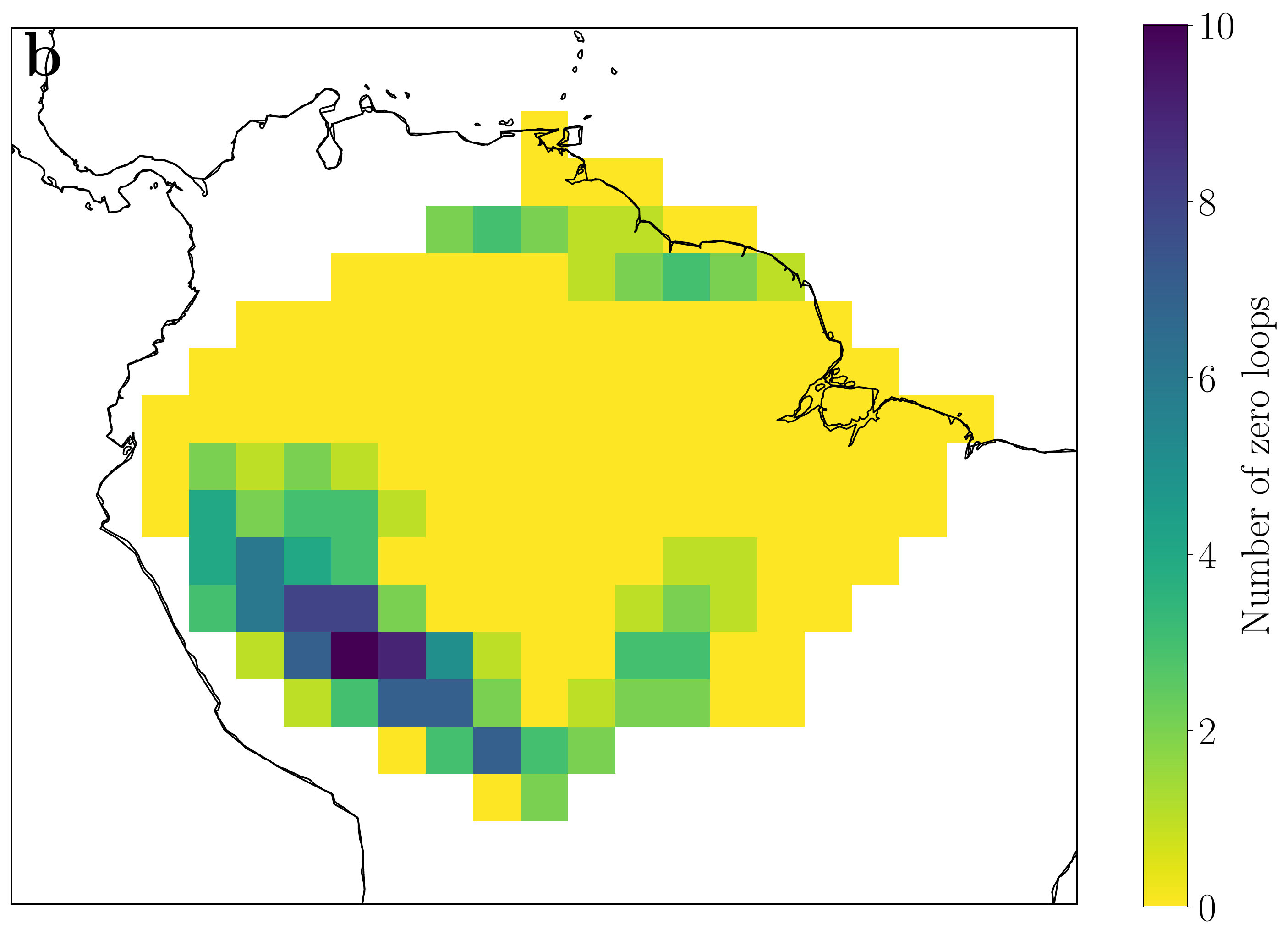}}
\subfigure{\includegraphics[width=.45\textwidth]{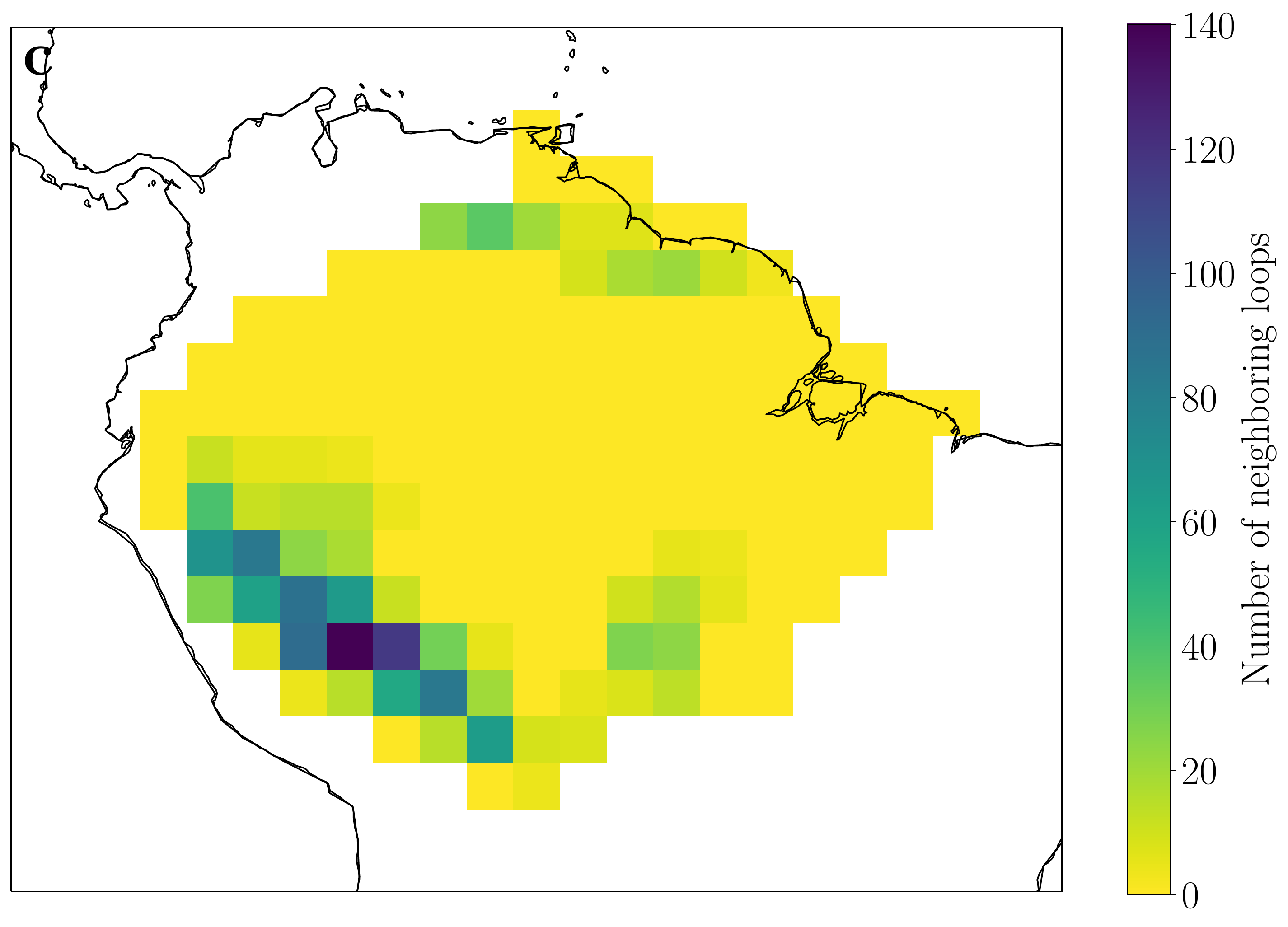}}
\subfigure{\includegraphics[width=.45\textwidth]{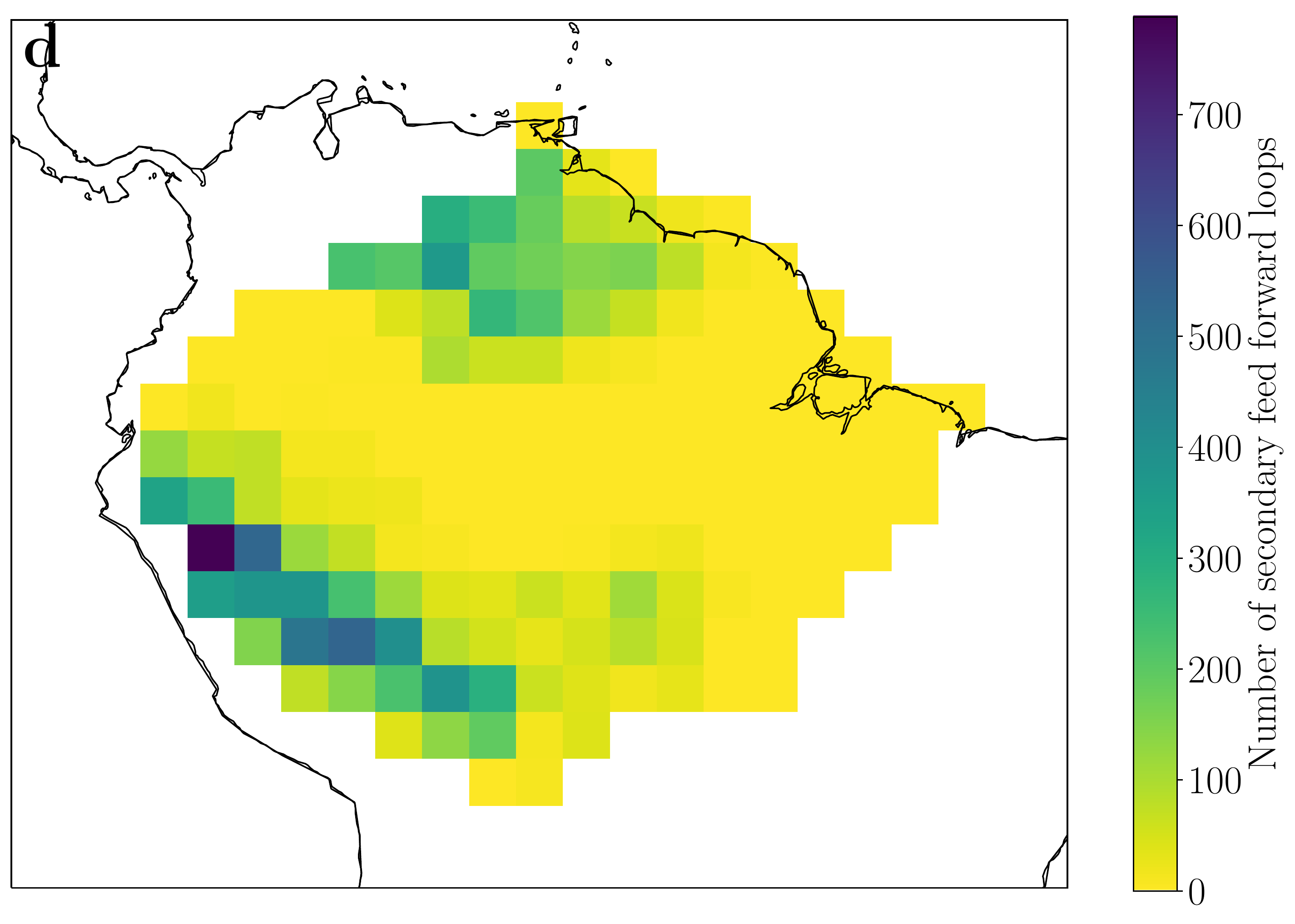}}
\caption{Number of motifs that point to a certain location in the 2x2$^\circ$ grid. Two regions with different vulnerability regimes can clearly be distinguished for all four investigated motifs. \textbf{a)} Feed forward loop, \textbf{b)} Zero loop, \textbf{c)} Neighboring loop, \textbf{d)} Secondary feed forward loop. The average degree is 5 and the number of nodes 160. Note the different colorbar scale between the sub-plots.}
\label{fig:six}
\end{figure}

Motifs also foster connectivity in real-world networks, for instance in medicine, food webs or the world wide web, carrying information forward~\citep{stouffer2012evolutionary, marinho2016authorship,alon2007network,lahav2004dynamics}. Basically, each network consists of certain motif structures that might be essential for the dynamics of the whole graph. One such example could be the moisture recycling network of the Amazon rainforest. It has been proposed that the Amazon rainforest is a tipping element with respect to the local precipitation~\citep{Lenton2008,nobre2016land}, which is suggested by conceptual models~\citep{van2014tipping} and data suggesting multistability of the rainforest~\citep{hirota2011global,staal2018forest,staal2015synergistic}, also on the regional scale.\\
Here, we construct a moisture recycling network in the Amazon rainforest and use the moisture flow data from tree transpiration on a 2$\times$2$^\circ$ resolution over the Amazon basin for the year 2014. The data has been created in Staal et al.~(2018)~\citep{staal2018forest}. Each node represents a 2$\times$2$^\circ$ patch of the rainforest and each link represents the atmospheric moisture transport from forest transpiration from one cell to another. To be able to compare the moisture recycling network with random networks, we construct the network in such a way that the average degree is the same as for the Erd\H{o}s-Rényi case. If we want to, for instance, achieve an average degree of 5, we only set the 160$\times$5 = 800 strongest moisture transport links between two nodes are regarded. Since this procedure favors strong connections, some weaker teleconnections between grid cells that are further away are lost. However, the dominant links remain such that the main network topology is preserved.\\
The coupling strength of these links is then set to the same value and the remaining connections are set to zero. Other effects are also neglected in this network since the aim is here to focus on the local and regional microstructures of the moisture recycling network and making it comparable to random networks. With these simplifications, we intend to investigate the structure and the possible implications it could have, instead of realistically modeling the tipping behavior of the Amazon rainforest. Similar approaches on viewing the Amazon rainforest as a complex network have been used earlier in literature~\citep{zemp2014importance,zemp2017self}. In these studies, it is shown that forest loss might be self-amplified in the Amazon basin if moisture recycling in the network is reduced, e.g., due to deforestation, and might lead to adverse cascading effects.\\
We evaluate the critical coupling that is necessary to start a cascade comparing the occurrence of tipping cascades between random networks and the moisture recycling network in the Amazon rainforest dependent on the coupling strength (Fig.~\ref{fig:five}; compare with Fig.~\ref{fig:two}). We reveal jumps in occurrence of cascade effects in the moisture recycling network when the coupling exceeds the critical strength of the feed forward loop. This is already the case for very sparse networks at low average degrees which hints at a highly clustered network with very localized motif structures (see also Fig.~\ref{fig:six}). Due to this structure, the moisture recycling network shows significantly more cascade effects at coupling strengths below 0.183 (single coupling) for low average degrees. For the other, weaker motifs, a step-like structure in the tipping cascades of the Amazon rainforest can hardly be noticed. Thus, these motifs only play a minor role in comparison to those in Erd\H{o}s-Rényi networks. The highly clustered moisture recycling network facilitates the likelihood for combinations of micro motifs that significantly elevate tipping cascades at lower couplings than in random networks. Hence, our results provide additional evidence that the Amazon network is more vulnerable than random networks following up on other aspects investigated in an earlier study~\citep{kronke2019dynamics}. For the same reason, more cascades occur in random networks than in the moisture recycling network for high coupling strengths (greater than 0.183) at the same average degree, since some parts of the Amazon rainforest network remain unconnected, because links between the closely connected clusters of highly connected areas are rare. 

In the remainder of this section, we show results for an average degree 5, but the results are robust against other average degrees (see supp. Fig.~S4). Basic motifs in the Amazon rainforest occur approximately 10 times more often than in the random network (Table~\ref{tab:compare}). This is due to the high connectivity in the Amazon rainforest in some regions of the network, while others are barely connected at all (Fig.~\ref{fig:six}). The clustering coefficient also hints at this network property of the Amazon network since it is one magnitude higher than in the Erd\H{o}s-Rényi network (0.297 versus 0.031$\pm$0.001). An overexpression, especially of the feed forward loop, has also been found in other real-world networks in biology or technology~\cite{Milo2002} suggesting an enhanced information or material flow through this network structure.

\begin{table}[htbp]
\centering
\begin{tabular}{l|c|c}
Number of motif occurrence & Amazon rainforest & Erd\H{o}s-Rényi \\ \hline
Feed forward loop & 2378 & 123$\pm$2\\
Zero loop & 168 & 25$\pm$1\\
Neighboring loop & 1499 & 149$\pm$6\\
Secondary feed forward loop & 11831 & 723$\pm$13\\
\\
\end{tabular}
\caption{Comparison of occurrence of a motif between the Amazon rainforest network with an average degree of 5 (and network size 160) with the respective Erd\H{o}s-Rényi networks, both with network size 160 and average degree 5. The clustering coefficient of the Erd\H{o}s-Rényi network is 0.031$\pm$0.001 and 0.297 for the Amazon network.
}
\label{tab:compare}
\end{table}

In Fig.~\ref{fig:six}, the occurrence of the four motifs are mapped. There are two major regions where motifs occur more frequently than in others. The first major region is located in the north of the rainforest, and the second in the south-west. These regions with a high occurrence of motifs, most importantly the feed forward loop, rely more on tree transpiration than other parts of the rainforest. However, even though the moisture transport link strength varies from connection to connection in our network (from 10 to a bit more than 100 millimeters per year), the number of motifs, especially for the feed forward loop, indicates a reduced stability against tipping cascades. Thus, it can be expected that these regions are more vulnerable than others in terms of changing rainfall conditions such that potential cascades could emerge faster.

\section{Discussion \& Conclusion}
In this study, we found that network motifs are able to disseminate critical transitions of tipping elements to further network nodes and can thus foster the emergence of tipping cascades. We worked out how motifs decrease the critical coupling strength that is necessary to start a cascade and quantified the occurrence of simple, but decisive micro structures. We detected that feed forward loops, the strongest three-node motif in our study, occur in sparse networks thus suggesting the existence of important hub nodes that tend to be more vulnerable than others and are prone to start a cascade. This seems to be of special importance also for real-world networks since there feed forward loops are often significantly overexpressed which has been found in the Amazon moisture recycling network example here. Similar findings have been made in other systems reported in literature~\citep{Milo2002}. There it is also stated that six types of feed forward loops exist which are combinations of the motifs described in this paper, i.e., one or more zero loops on top of a feed forward loop. These specific combinations could be interesting to be investigated in future research in more detail due to the importance of the feed forward loop motif. However, we restricted our analysis to the four fundamental motifs since it is their fingerprint that is observed in the tipping cascade experiments (see Fig.~\ref{fig:two}).
Additionally, we find in our experiments that the critical coupling strength in densely connected Erd\H{o}s-Rényi networks is more than 90\% lower compared to the single coupling of two tipping elements, due to coupled feed forward loops. Thus, we are able to understand the occurrence of tipping cascades in sparse and dense random networks.\\
In the Amazon rainforest application, the location of motifs is highly clustered in distinct areas of the rainforest thus indicating increased vulnerability in these locations. There, tipping cascades can emerge at lower couplings than they could for Erd\H{o}s-Rényi networks. In turn, this would also imply that reforestation in these regions is more effective. \\
However, this conclusion is limited by the simplified nature of the Amazon network realization in this study and could be examined further by the usage of the actual moisture recycling values in a more in-depth study. Since the model employed here is simplified and conceptual, the question remains if the role of motifs would change under more realistic model realisations or other dynamics on the nodes themselves.\\
In turn, it might be insightful to investigate the role of motifs on other fully dynamic systems that are connected via a network structure, for instance in food webs, transcriptional networks or Earth system components.

\newpage
\section*{Supplementary Material}
See supplementary material for more details on the scaling for the weaker motifs (zero loop, neighboring loop and secondary feed forward loop) in parallel to the feed forward loop (Fig.~\ref{fig:three}b). Furthermore, details on multiple motifs and the degree dependency of the feed forward loop in the moisture recycling network are shown.

\begin{acknowledgments}
This work has been carried out within the framework of the PIK FutureLab on Earth Resilience in the Anthropocene. N.W. and R.W. acknowledge the financial support of the IRTG 1740/TRP 2015/50122-0 project funded by DFG and FAPESP. N.W. is grateful for a scholarship from the Studienstiftung des deutschen Volkes. A.S. and J.F.D. acknowledge support from the European Research Council project Earth Resilience in the Anthropocene (743080 ERA). A.S. and O.A.T. thank for support from the Bolin Centre for Climate Research. O.A.T. acknowledges funding from the Netherlands Organisation for Scientific Research Innovational Research Incentives Schemes VENI (016.171.019). J.F.D. is grateful for financial support by the Stordalen Foundation via the Planetary Boundary Research Network (PB.net) and the Earth League’s EarthDoc program. We are thankful for financial support by the Leibniz Association (project DominoES). The authors gratefully acknowledge the European Regional Development Fund (ERDF), the German Federal Ministry of Education and Research and the Land Brandenburg for supporting this project by providing resources on the high performance computer system at the Potsdam Institute for Climate Impact Research. The authors also gratefully acknowledge discussions with Ann-Kristin Klose, Marc Wiedermann and Jobst Heitzig.\\
\end{acknowledgments}


\textbf{Data availability}\\
The data that was used in this study is available from the corresponding author upon reasonable request.\\

\textbf{Competing financial interests}\\
The authors declare no competing financial interests.

\nocite{*}
\bibliography{aipsamp}

\providecommand{\noopsort}[1]{}\providecommand{\singleletter}[1]{#1}%
\begin{thebibliography}{44}%
\makeatletter
\providecommand \@ifxundefined [1]{%
 \@ifx{#1\undefined}
}%
\providecommand \@ifnum [1]{%
 \ifnum #1\expandafter \@firstoftwo
 \else \expandafter \@secondoftwo
 \fi
}%
\providecommand \@ifx [1]{%
 \ifx #1\expandafter \@firstoftwo
 \else \expandafter \@secondoftwo
 \fi
}%
\providecommand \natexlab [1]{#1}%
\providecommand \enquote  [1]{``#1''}%
\providecommand \bibnamefont  [1]{#1}%
\providecommand \bibfnamefont [1]{#1}%
\providecommand \citenamefont [1]{#1}%
\providecommand \href@noop [0]{\@secondoftwo}%
\providecommand \href [0]{\begingroup \@sanitize@url \@href}%
\providecommand \@href[1]{\@@startlink{#1}\@@href}%
\providecommand \@@href[1]{\endgroup#1\@@endlink}%
\providecommand \@sanitize@url [0]{\catcode `\\12\catcode `\$12\catcode
  `\&12\catcode `\#12\catcode `\^12\catcode `\_12\catcode `\%12\relax}%
\providecommand \@@startlink[1]{}%
\providecommand \@@endlink[0]{}%
\providecommand \url  [0]{\begingroup\@sanitize@url \@url }%
\providecommand \@url [1]{\endgroup\@href {#1}{\urlprefix }}%
\providecommand \urlprefix  [0]{URL }%
\providecommand \Eprint [0]{\href }%
\providecommand \doibase [0]{http://dx.doi.org/}%
\providecommand \selectlanguage [0]{\@gobble}%
\providecommand \bibinfo  [0]{\@secondoftwo}%
\providecommand \bibfield  [0]{\@secondoftwo}%
\providecommand \translation [1]{[#1]}%
\providecommand \BibitemOpen [0]{}%
\providecommand \bibitemStop [0]{}%
\providecommand \bibitemNoStop [0]{.\EOS\space}%
\providecommand \EOS [0]{\spacefactor3000\relax}%
\providecommand \BibitemShut  [1]{\csname bibitem#1\endcsname}%
\let\auto@bib@innerbib\@empty
\bibitem [{\citenamefont {Newman}(2003)}]{newman2003structure}%
  \BibitemOpen
  \bibfield  {author} {\bibinfo {author} {\bibfnamefont {M.~E.}\ \bibnamefont
  {Newman}},\ }\bibfield  {title} {\enquote {\bibinfo {title} {The structure
  and function of complex networks},}\ }\href@noop {} {\bibfield  {journal}
  {\bibinfo  {journal} {SIAM review}\ }\textbf {\bibinfo {volume} {45}},\
  \bibinfo {pages} {167--256} (\bibinfo {year} {2003})}\BibitemShut {NoStop}%
\bibitem [{\citenamefont {Zou}\ \emph {et~al.}(2013)\citenamefont {Zou},
  \citenamefont {Senthilkumar}, \citenamefont {Zhan},\ and\ \citenamefont
  {Kurths}}]{zou2013reviving}%
  \BibitemOpen
  \bibfield  {author} {\bibinfo {author} {\bibfnamefont {W.}~\bibnamefont
  {Zou}}, \bibinfo {author} {\bibfnamefont {D.}~\bibnamefont {Senthilkumar}},
  \bibinfo {author} {\bibfnamefont {M.}~\bibnamefont {Zhan}}, \ and\ \bibinfo
  {author} {\bibfnamefont {J.}~\bibnamefont {Kurths}},\ }\bibfield  {title}
  {\enquote {\bibinfo {title} {Reviving oscillations in coupled nonlinear
  oscillators},}\ }\href@noop {} {\bibfield  {journal} {\bibinfo  {journal}
  {Physical Review Letters}\ }\textbf {\bibinfo {volume} {111}},\ \bibinfo
  {pages} {014101} (\bibinfo {year} {2013})}\BibitemShut {NoStop}%
\bibitem [{\citenamefont {Gross}\ \emph {et~al.}(2009)\citenamefont {Gross},
  \citenamefont {Rudolf}, \citenamefont {Levin},\ and\ \citenamefont
  {Dieckmann}}]{gross2009generalized}%
  \BibitemOpen
  \bibfield  {author} {\bibinfo {author} {\bibfnamefont {T.}~\bibnamefont
  {Gross}}, \bibinfo {author} {\bibfnamefont {L.}~\bibnamefont {Rudolf}},
  \bibinfo {author} {\bibfnamefont {S.~A.}\ \bibnamefont {Levin}}, \ and\
  \bibinfo {author} {\bibfnamefont {U.}~\bibnamefont {Dieckmann}},\ }\bibfield
  {title} {\enquote {\bibinfo {title} {Generalized models reveal stabilizing
  factors in food webs},}\ }\href@noop {} {\bibfield  {journal} {\bibinfo
  {journal} {Science}\ }\textbf {\bibinfo {volume} {325}},\ \bibinfo {pages}
  {747--750} (\bibinfo {year} {2009})}\BibitemShut {NoStop}%
\bibitem [{\citenamefont {Nitzbon}\ \emph {et~al.}(2017)\citenamefont
  {Nitzbon}, \citenamefont {Schultz}, \citenamefont {Heitzig}, \citenamefont
  {Kurths},\ and\ \citenamefont {Hellmann}}]{nitzbon2017deciphering}%
  \BibitemOpen
  \bibfield  {author} {\bibinfo {author} {\bibfnamefont {J.}~\bibnamefont
  {Nitzbon}}, \bibinfo {author} {\bibfnamefont {P.}~\bibnamefont {Schultz}},
  \bibinfo {author} {\bibfnamefont {J.}~\bibnamefont {Heitzig}}, \bibinfo
  {author} {\bibfnamefont {J.}~\bibnamefont {Kurths}}, \ and\ \bibinfo {author}
  {\bibfnamefont {F.}~\bibnamefont {Hellmann}},\ }\bibfield  {title} {\enquote
  {\bibinfo {title} {Deciphering the imprint of topology on nonlinear dynamical
  network stability},}\ }\href@noop {} {\bibfield  {journal} {\bibinfo
  {journal} {New Journal of Physics}\ }\textbf {\bibinfo {volume} {19}},\
  \bibinfo {pages} {033029} (\bibinfo {year} {2017})}\BibitemShut {NoStop}%
\bibitem [{\citenamefont {Lenton}\ \emph {et~al.}(2008)\citenamefont {Lenton},
  \citenamefont {Held}, \citenamefont {Kriegler}, \citenamefont {Hall},
  \citenamefont {Lucht}, \citenamefont {Rahmstorf},\ and\ \citenamefont
  {Schellnhuber}}]{Lenton2008}%
  \BibitemOpen
  \bibfield  {author} {\bibinfo {author} {\bibfnamefont {T.~M.}\ \bibnamefont
  {Lenton}}, \bibinfo {author} {\bibfnamefont {H.}~\bibnamefont {Held}},
  \bibinfo {author} {\bibfnamefont {E.}~\bibnamefont {Kriegler}}, \bibinfo
  {author} {\bibfnamefont {J.~W.}\ \bibnamefont {Hall}}, \bibinfo {author}
  {\bibfnamefont {W.}~\bibnamefont {Lucht}}, \bibinfo {author} {\bibfnamefont
  {S.}~\bibnamefont {Rahmstorf}}, \ and\ \bibinfo {author} {\bibfnamefont
  {H.~J.}\ \bibnamefont {Schellnhuber}},\ }\bibfield  {title} {\enquote
  {\bibinfo {title} {Tipping elements in the {E}arth{\textquoteright}s climate
  system},}\ }\href {\doibase 10.1073/pnas.0705414105} {\bibfield  {journal}
  {\bibinfo  {journal} {Proceedings of the National Academy of Sciences}\
  }\textbf {\bibinfo {volume} {105}},\ \bibinfo {pages} {1786--1793} (\bibinfo
  {year} {2008})}\BibitemShut {NoStop}%
\bibitem [{\citenamefont {Brummitt}, \citenamefont {Barnett},\ and\
  \citenamefont {D'Souza}(2015)}]{Brummit2015}%
  \BibitemOpen
  \bibfield  {author} {\bibinfo {author} {\bibfnamefont {C.~D.}\ \bibnamefont
  {Brummitt}}, \bibinfo {author} {\bibfnamefont {G.}~\bibnamefont {Barnett}}, \
  and\ \bibinfo {author} {\bibfnamefont {R.~M.}\ \bibnamefont {D'Souza}},\
  }\bibfield  {title} {\enquote {\bibinfo {title} {Coupled catastrophes: sudden
  shifts cascade and hop among interdependent systems},}\ }\href@noop {}
  {\bibfield  {journal} {\bibinfo  {journal} {J. Royal Soc. Interface}\
  }\textbf {\bibinfo {volume} {12}},\ \bibinfo {pages} {20150712} (\bibinfo
  {year} {2015})}\BibitemShut {NoStop}%
\bibitem [{\citenamefont {Kriegler}\ \emph {et~al.}(2009)\citenamefont
  {Kriegler}, \citenamefont {Hall}, \citenamefont {Held}, \citenamefont
  {Dawson},\ and\ \citenamefont {Schellnhuber}}]{Kriegler2009}%
  \BibitemOpen
  \bibfield  {author} {\bibinfo {author} {\bibfnamefont {E.}~\bibnamefont
  {Kriegler}}, \bibinfo {author} {\bibfnamefont {J.~W.}\ \bibnamefont {Hall}},
  \bibinfo {author} {\bibfnamefont {H.}~\bibnamefont {Held}}, \bibinfo {author}
  {\bibfnamefont {R.}~\bibnamefont {Dawson}}, \ and\ \bibinfo {author}
  {\bibfnamefont {H.~J.}\ \bibnamefont {Schellnhuber}},\ }\bibfield  {title}
  {\enquote {\bibinfo {title} {Imprecise probability assessment of tipping
  points in the climate system},}\ }\href {\doibase 10.1073/pnas.0809117106}
  {\bibfield  {journal} {\bibinfo  {journal} {Proceedings of the National
  Academy of Sciences}\ }\textbf {\bibinfo {volume} {106}},\ \bibinfo {pages}
  {5041--5046} (\bibinfo {year} {2009})}\BibitemShut {NoStop}%
\bibitem [{\citenamefont {Cai}, \citenamefont {Lenton},\ and\ \citenamefont
  {Lontzek}(2016)}]{Cai2016}%
  \BibitemOpen
  \bibfield  {author} {\bibinfo {author} {\bibfnamefont {Y.}~\bibnamefont
  {Cai}}, \bibinfo {author} {\bibfnamefont {T.~M.}\ \bibnamefont {Lenton}}, \
  and\ \bibinfo {author} {\bibfnamefont {T.~S.}\ \bibnamefont {Lontzek}},\
  }\bibfield  {title} {\enquote {\bibinfo {title} {Risk of multiple interacting
  tipping points should encourage rapid {C}{O}{2} emission reduction},}\ }\href
  {\doibase 10.1038/nclimate2964} {\bibfield  {journal} {\bibinfo  {journal}
  {Nat. Clim. Chang.}\ }\textbf {\bibinfo {volume} {6}},\ \bibinfo {pages}
  {520--525} (\bibinfo {year} {2016})}\BibitemShut {NoStop}%
\bibitem [{\citenamefont {Rocha}\ \emph {et~al.}(2018)\citenamefont {Rocha},
  \citenamefont {Peterson}, \citenamefont {Bodin},\ and\ \citenamefont
  {Levin}}]{rocha2018cascading}%
  \BibitemOpen
  \bibfield  {author} {\bibinfo {author} {\bibfnamefont {J.~C.}\ \bibnamefont
  {Rocha}}, \bibinfo {author} {\bibfnamefont {G.}~\bibnamefont {Peterson}},
  \bibinfo {author} {\bibfnamefont {{\"O}.}~\bibnamefont {Bodin}}, \ and\
  \bibinfo {author} {\bibfnamefont {S.}~\bibnamefont {Levin}},\ }\bibfield
  {title} {\enquote {\bibinfo {title} {Cascading regime shifts within and
  across scales},}\ }\href@noop {} {\bibfield  {journal} {\bibinfo  {journal}
  {Science}\ }\textbf {\bibinfo {volume} {362}},\ \bibinfo {pages} {1379--1383}
  (\bibinfo {year} {2018})}\BibitemShut {NoStop}%
\bibitem [{\citenamefont {Gaucherel}\ and\ \citenamefont
  {Moron}(2017)}]{gaucherel2017potential}%
  \BibitemOpen
  \bibfield  {author} {\bibinfo {author} {\bibfnamefont {C.}~\bibnamefont
  {Gaucherel}}\ and\ \bibinfo {author} {\bibfnamefont {V.}~\bibnamefont
  {Moron}},\ }\bibfield  {title} {\enquote {\bibinfo {title} {Potential
  stabilizing points to mitigate tipping point interactions in earth's
  climate},}\ }\href@noop {} {\bibfield  {journal} {\bibinfo  {journal}
  {International Journal of Climatology}\ }\textbf {\bibinfo {volume} {37}},\
  \bibinfo {pages} {399--408} (\bibinfo {year} {2017})}\BibitemShut {NoStop}%
\bibitem [{\citenamefont {Dekker}, \citenamefont {Heydt},\ and\ \citenamefont
  {Dijkstra}(2018)}]{dekker2018cascading}%
  \BibitemOpen
  \bibfield  {author} {\bibinfo {author} {\bibfnamefont {M.~M.}\ \bibnamefont
  {Dekker}}, \bibinfo {author} {\bibfnamefont {A.~S.}\ \bibnamefont {Heydt}}, \
  and\ \bibinfo {author} {\bibfnamefont {H.~A.}\ \bibnamefont {Dijkstra}},\
  }\bibfield  {title} {\enquote {\bibinfo {title} {Cascading transitions in the
  climate system},}\ }\href@noop {} {\bibfield  {journal} {\bibinfo  {journal}
  {Earth System Dynamics}\ }\textbf {\bibinfo {volume} {9}},\ \bibinfo {pages}
  {1243--1260} (\bibinfo {year} {2018})}\BibitemShut {NoStop}%
\bibitem [{\citenamefont {Klose}\ \emph {et~al.}(2019)\citenamefont {Klose},
  \citenamefont {Karle}, \citenamefont {Winkelmann},\ and\ \citenamefont
  {Donges}}]{klose2019dynamic}%
  \BibitemOpen
  \bibfield  {author} {\bibinfo {author} {\bibfnamefont {A.~K.}\ \bibnamefont
  {Klose}}, \bibinfo {author} {\bibfnamefont {V.}~\bibnamefont {Karle}},
  \bibinfo {author} {\bibfnamefont {R.}~\bibnamefont {Winkelmann}}, \ and\
  \bibinfo {author} {\bibfnamefont {J.~F.}\ \bibnamefont {Donges}},\ }\bibfield
   {title} {\enquote {\bibinfo {title} {Dynamic emergence of domino effects in
  systems of interacting tipping elements in ecology and climate},}\
  }\href@noop {} {\bibfield  {journal} {\bibinfo  {journal} {arXiv preprint
  arXiv:1910.12042}\ } (\bibinfo {year} {2019})}\BibitemShut {NoStop}%
\bibitem [{\citenamefont {Steffen}\ \emph {et~al.}(2018)\citenamefont
  {Steffen}, \citenamefont {Rockstr{\"o}m}, \citenamefont {Richardson},
  \citenamefont {Lenton}, \citenamefont {Folke}, \citenamefont {Liverman},
  \citenamefont {Summerhayes}, \citenamefont {Barnosky}, \citenamefont
  {Cornell}, \citenamefont {Crucifix} \emph
  {et~al.}}]{steffen2018trajectories}%
  \BibitemOpen
  \bibfield  {author} {\bibinfo {author} {\bibfnamefont {W.}~\bibnamefont
  {Steffen}}, \bibinfo {author} {\bibfnamefont {J.}~\bibnamefont
  {Rockstr{\"o}m}}, \bibinfo {author} {\bibfnamefont {K.}~\bibnamefont
  {Richardson}}, \bibinfo {author} {\bibfnamefont {T.~M.}\ \bibnamefont
  {Lenton}}, \bibinfo {author} {\bibfnamefont {C.}~\bibnamefont {Folke}},
  \bibinfo {author} {\bibfnamefont {D.}~\bibnamefont {Liverman}}, \bibinfo
  {author} {\bibfnamefont {C.~P.}\ \bibnamefont {Summerhayes}}, \bibinfo
  {author} {\bibfnamefont {A.~D.}\ \bibnamefont {Barnosky}}, \bibinfo {author}
  {\bibfnamefont {S.~E.}\ \bibnamefont {Cornell}}, \bibinfo {author}
  {\bibfnamefont {M.}~\bibnamefont {Crucifix}},  \emph {et~al.},\ }\bibfield
  {title} {\enquote {\bibinfo {title} {Trajectories of the {E}arth {S}ystem in
  the {A}nthropocene},}\ }\href@noop {} {\bibfield  {journal} {\bibinfo
  {journal} {Proceedings of the National Academy of Sciences}\ }\textbf
  {\bibinfo {volume} {115}},\ \bibinfo {pages} {8252--8259} (\bibinfo {year}
  {2018})}\BibitemShut {NoStop}%
\bibitem [{\citenamefont {Kr{\"o}nke}\ \emph {et~al.}(2019)\citenamefont
  {Kr{\"o}nke}, \citenamefont {Wunderling}, \citenamefont {Winkelmann},
  \citenamefont {Staal}, \citenamefont {Stumpf}, \citenamefont {Tuinenburg},\
  and\ \citenamefont {Donges}}]{kronke2019dynamics}%
  \BibitemOpen
  \bibfield  {author} {\bibinfo {author} {\bibfnamefont {J.}~\bibnamefont
  {Kr{\"o}nke}}, \bibinfo {author} {\bibfnamefont {N.}~\bibnamefont
  {Wunderling}}, \bibinfo {author} {\bibfnamefont {R.}~\bibnamefont
  {Winkelmann}}, \bibinfo {author} {\bibfnamefont {A.}~\bibnamefont {Staal}},
  \bibinfo {author} {\bibfnamefont {B.}~\bibnamefont {Stumpf}}, \bibinfo
  {author} {\bibfnamefont {O.~A.}\ \bibnamefont {Tuinenburg}}, \ and\ \bibinfo
  {author} {\bibfnamefont {J.~F.}\ \bibnamefont {Donges}},\ }\bibfield  {title}
  {\enquote {\bibinfo {title} {Dynamics of {T}ipping {C}ascades on {C}omplex
  {N}etworks},}\ }\href@noop {} {\bibfield  {journal} {\bibinfo  {journal}
  {arXiv preprint arXiv:1905.05476}\ } (\bibinfo {year} {2019})}\BibitemShut
  {NoStop}%
\bibitem [{\citenamefont {Eom}(2018)}]{eom2018resilience}%
  \BibitemOpen
  \bibfield  {author} {\bibinfo {author} {\bibfnamefont {Y.-H.}\ \bibnamefont
  {Eom}},\ }\bibfield  {title} {\enquote {\bibinfo {title} {Resilience of
  networks to environmental stress: From regular to random networks},}\
  }\href@noop {} {\bibfield  {journal} {\bibinfo  {journal} {Physical Review
  E}\ }\textbf {\bibinfo {volume} {97}},\ \bibinfo {pages} {042313} (\bibinfo
  {year} {2018})}\BibitemShut {NoStop}%
\bibitem [{\citenamefont {Watts}(2002)}]{watts2002simple}%
  \BibitemOpen
  \bibfield  {author} {\bibinfo {author} {\bibfnamefont {D.~J.}\ \bibnamefont
  {Watts}},\ }\bibfield  {title} {\enquote {\bibinfo {title} {A simple model of
  global cascades on random networks},}\ }\href@noop {} {\bibfield  {journal}
  {\bibinfo  {journal} {Proceedings of the National Academy of Sciences}\
  }\textbf {\bibinfo {volume} {99}},\ \bibinfo {pages} {5766--5771} (\bibinfo
  {year} {2002})}\BibitemShut {NoStop}%
\bibitem [{\citenamefont {Buldyrev}\ \emph {et~al.}(2010)\citenamefont
  {Buldyrev}, \citenamefont {Parshani}, \citenamefont {Paul}, \citenamefont
  {Stanley},\ and\ \citenamefont {Havlin}}]{buldyrev2010catastrophic}%
  \BibitemOpen
  \bibfield  {author} {\bibinfo {author} {\bibfnamefont {S.~V.}\ \bibnamefont
  {Buldyrev}}, \bibinfo {author} {\bibfnamefont {R.}~\bibnamefont {Parshani}},
  \bibinfo {author} {\bibfnamefont {G.}~\bibnamefont {Paul}}, \bibinfo {author}
  {\bibfnamefont {H.~E.}\ \bibnamefont {Stanley}}, \ and\ \bibinfo {author}
  {\bibfnamefont {S.}~\bibnamefont {Havlin}},\ }\bibfield  {title} {\enquote
  {\bibinfo {title} {Catastrophic cascade of failures in interdependent
  networks},}\ }\href@noop {} {\bibfield  {journal} {\bibinfo  {journal}
  {Nature}\ }\textbf {\bibinfo {volume} {464}},\ \bibinfo {pages} {1025}
  (\bibinfo {year} {2010})}\BibitemShut {NoStop}%
\bibitem [{\citenamefont {Turalska}\ \emph {et~al.}(2019)\citenamefont
  {Turalska}, \citenamefont {Burghardt}, \citenamefont {Rohden}, \citenamefont
  {Swami},\ and\ \citenamefont {D'Souza}}]{turalska2019cascading}%
  \BibitemOpen
  \bibfield  {author} {\bibinfo {author} {\bibfnamefont {M.}~\bibnamefont
  {Turalska}}, \bibinfo {author} {\bibfnamefont {K.}~\bibnamefont {Burghardt}},
  \bibinfo {author} {\bibfnamefont {M.}~\bibnamefont {Rohden}}, \bibinfo
  {author} {\bibfnamefont {A.}~\bibnamefont {Swami}}, \ and\ \bibinfo {author}
  {\bibfnamefont {R.~M.}\ \bibnamefont {D'Souza}},\ }\bibfield  {title}
  {\enquote {\bibinfo {title} {Cascading failures in scale-free interdependent
  networks},}\ }\href@noop {} {\bibfield  {journal} {\bibinfo  {journal}
  {Physical Review E}\ }\textbf {\bibinfo {volume} {99}},\ \bibinfo {pages}
  {032308} (\bibinfo {year} {2019})}\BibitemShut {NoStop}%
\bibitem [{\citenamefont {Loppini}, \citenamefont {Filippi},\ and\
  \citenamefont {Stanley}(2019)}]{loppini2019critical}%
  \BibitemOpen
  \bibfield  {author} {\bibinfo {author} {\bibfnamefont {A.}~\bibnamefont
  {Loppini}}, \bibinfo {author} {\bibfnamefont {S.}~\bibnamefont {Filippi}}, \
  and\ \bibinfo {author} {\bibfnamefont {H.~E.}\ \bibnamefont {Stanley}},\
  }\bibfield  {title} {\enquote {\bibinfo {title} {Critical transitions in
  heterogeneous networks: Loss of low-degree nodes as an early warning
  signal},}\ }\href@noop {} {\bibfield  {journal} {\bibinfo  {journal}
  {Physical Review E}\ }\textbf {\bibinfo {volume} {99}},\ \bibinfo {pages}
  {040301} (\bibinfo {year} {2019})}\BibitemShut {NoStop}%
\bibitem [{\citenamefont {Wu}\ \emph {et~al.}(2018)\citenamefont {Wu},
  \citenamefont {Fennell}, \citenamefont {Percus}, \citenamefont {Lerman} \emph
  {et~al.}}]{wu2018degree}%
  \BibitemOpen
  \bibfield  {author} {\bibinfo {author} {\bibfnamefont {X.-Z.}\ \bibnamefont
  {Wu}}, \bibinfo {author} {\bibfnamefont {P.~G.}\ \bibnamefont {Fennell}},
  \bibinfo {author} {\bibfnamefont {A.~G.}\ \bibnamefont {Percus}}, \bibinfo
  {author} {\bibfnamefont {K.}~\bibnamefont {Lerman}},  \emph {et~al.},\
  }\bibfield  {title} {\enquote {\bibinfo {title} {Degree correlations amplify
  the growth of cascades in networks},}\ }\href@noop {} {\bibfield  {journal}
  {\bibinfo  {journal} {Physical Review E}\ }\textbf {\bibinfo {volume} {98}},\
  \bibinfo {pages} {022321} (\bibinfo {year} {2018})}\BibitemShut {NoStop}%
\bibitem [{\citenamefont {Liu}\ \emph {et~al.}(2019)\citenamefont {Liu},
  \citenamefont {Pan}, \citenamefont {Stanley},\ and\ \citenamefont
  {Gao}}]{liu2019multiple}%
  \BibitemOpen
  \bibfield  {author} {\bibinfo {author} {\bibfnamefont {X.}~\bibnamefont
  {Liu}}, \bibinfo {author} {\bibfnamefont {L.}~\bibnamefont {Pan}}, \bibinfo
  {author} {\bibfnamefont {H.~E.}\ \bibnamefont {Stanley}}, \ and\ \bibinfo
  {author} {\bibfnamefont {J.}~\bibnamefont {Gao}},\ }\bibfield  {title}
  {\enquote {\bibinfo {title} {Multiple phase transitions in networks of
  directed networks},}\ }\href@noop {} {\bibfield  {journal} {\bibinfo
  {journal} {Physical Review E}\ }\textbf {\bibinfo {volume} {99}},\ \bibinfo
  {pages} {012312} (\bibinfo {year} {2019})}\BibitemShut {NoStop}%
\bibitem [{\citenamefont {Krishnagopal}\ \emph {et~al.}(2017)\citenamefont
  {Krishnagopal}, \citenamefont {Lehnert}, \citenamefont {Poel}, \citenamefont
  {Zakharova},\ and\ \citenamefont
  {Sch{\"o}ll}}]{krishnagopal2017synchronization}%
  \BibitemOpen
  \bibfield  {author} {\bibinfo {author} {\bibfnamefont {S.}~\bibnamefont
  {Krishnagopal}}, \bibinfo {author} {\bibfnamefont {J.}~\bibnamefont
  {Lehnert}}, \bibinfo {author} {\bibfnamefont {W.}~\bibnamefont {Poel}},
  \bibinfo {author} {\bibfnamefont {A.}~\bibnamefont {Zakharova}}, \ and\
  \bibinfo {author} {\bibfnamefont {E.}~\bibnamefont {Sch{\"o}ll}},\ }\bibfield
   {title} {\enquote {\bibinfo {title} {Synchronization patterns: from network
  motifs to hierarchical networks},}\ }\href@noop {} {\bibfield  {journal}
  {\bibinfo  {journal} {Philosophical Transactions of the Royal Society A:
  Mathematical, Physical and Engineering Sciences}\ }\textbf {\bibinfo {volume}
  {375}},\ \bibinfo {pages} {20160216} (\bibinfo {year} {2017})}\BibitemShut
  {NoStop}%
\bibitem [{\citenamefont {D’Huys}\ \emph {et~al.}(2008)\citenamefont
  {D’Huys}, \citenamefont {Vicente}, \citenamefont {Erneux}, \citenamefont
  {Danckaert},\ and\ \citenamefont {Fischer}}]{d2008synchronization}%
  \BibitemOpen
  \bibfield  {author} {\bibinfo {author} {\bibfnamefont {O.}~\bibnamefont
  {D’Huys}}, \bibinfo {author} {\bibfnamefont {R.}~\bibnamefont {Vicente}},
  \bibinfo {author} {\bibfnamefont {T.}~\bibnamefont {Erneux}}, \bibinfo
  {author} {\bibfnamefont {J.}~\bibnamefont {Danckaert}}, \ and\ \bibinfo
  {author} {\bibfnamefont {I.}~\bibnamefont {Fischer}},\ }\bibfield  {title}
  {\enquote {\bibinfo {title} {Synchronization properties of network motifs:
  Influence of coupling delay and symmetry},}\ }\href@noop {} {\bibfield
  {journal} {\bibinfo  {journal} {Chaos: An Interdisciplinary Journal of
  Nonlinear Science}\ }\textbf {\bibinfo {volume} {18}},\ \bibinfo {pages}
  {037116} (\bibinfo {year} {2008})}\BibitemShut {NoStop}%
\bibitem [{\citenamefont {Gambuzza}, \citenamefont {G{\'o}mez-Garde{\~n}es},\
  and\ \citenamefont {Frasca}(2016)}]{gambuzza2016amplitude}%
  \BibitemOpen
  \bibfield  {author} {\bibinfo {author} {\bibfnamefont {L.~V.}\ \bibnamefont
  {Gambuzza}}, \bibinfo {author} {\bibfnamefont {J.}~\bibnamefont
  {G{\'o}mez-Garde{\~n}es}}, \ and\ \bibinfo {author} {\bibfnamefont
  {M.}~\bibnamefont {Frasca}},\ }\bibfield  {title} {\enquote {\bibinfo {title}
  {Amplitude dynamics favors synchronization in complex networks},}\
  }\href@noop {} {\bibfield  {journal} {\bibinfo  {journal} {Scientific
  reports}\ }\textbf {\bibinfo {volume} {6}},\ \bibinfo {pages} {24915}
  (\bibinfo {year} {2016})}\BibitemShut {NoStop}%
\bibitem [{\citenamefont {Milo}\ \emph {et~al.}(2002)\citenamefont {Milo},
  \citenamefont {Shen-Orr}, \citenamefont {Itzkovitz}, \citenamefont {Kashtan},
  \citenamefont {Chklovskii},\ and\ \citenamefont {Alon}}]{Milo2002}%
  \BibitemOpen
  \bibfield  {author} {\bibinfo {author} {\bibfnamefont {R.}~\bibnamefont
  {Milo}}, \bibinfo {author} {\bibfnamefont {S.}~\bibnamefont {Shen-Orr}},
  \bibinfo {author} {\bibfnamefont {S.}~\bibnamefont {Itzkovitz}}, \bibinfo
  {author} {\bibfnamefont {N.}~\bibnamefont {Kashtan}}, \bibinfo {author}
  {\bibfnamefont {D.}~\bibnamefont {Chklovskii}}, \ and\ \bibinfo {author}
  {\bibfnamefont {U.}~\bibnamefont {Alon}},\ }\bibfield  {title} {\enquote
  {\bibinfo {title} {Network motifs: Simple building blocks of complex
  networks},}\ }\href {\doibase 10.1126/science.298.5594.824} {\bibfield
  {journal} {\bibinfo  {journal} {Science}\ }\textbf {\bibinfo {volume}
  {298}},\ \bibinfo {pages} {824--827} (\bibinfo {year} {2002})}\BibitemShut
  {NoStop}%
\bibitem [{\citenamefont {Stouffer}\ \emph {et~al.}(2012)\citenamefont
  {Stouffer}, \citenamefont {Sales-Pardo}, \citenamefont {Sirer},\ and\
  \citenamefont {Bascompte}}]{stouffer2012evolutionary}%
  \BibitemOpen
  \bibfield  {author} {\bibinfo {author} {\bibfnamefont {D.~B.}\ \bibnamefont
  {Stouffer}}, \bibinfo {author} {\bibfnamefont {M.}~\bibnamefont
  {Sales-Pardo}}, \bibinfo {author} {\bibfnamefont {M.~I.}\ \bibnamefont
  {Sirer}}, \ and\ \bibinfo {author} {\bibfnamefont {J.}~\bibnamefont
  {Bascompte}},\ }\bibfield  {title} {\enquote {\bibinfo {title} {Evolutionary
  conservation of species’ roles in food webs},}\ }\href@noop {} {\bibfield
  {journal} {\bibinfo  {journal} {Science}\ }\textbf {\bibinfo {volume}
  {335}},\ \bibinfo {pages} {1489--1492} (\bibinfo {year} {2012})}\BibitemShut
  {NoStop}%
\bibitem [{\citenamefont {Marinho}, \citenamefont {Hirst},\ and\ \citenamefont
  {Amancio}(2016)}]{marinho2016authorship}%
  \BibitemOpen
  \bibfield  {author} {\bibinfo {author} {\bibfnamefont {V.~Q.}\ \bibnamefont
  {Marinho}}, \bibinfo {author} {\bibfnamefont {G.}~\bibnamefont {Hirst}}, \
  and\ \bibinfo {author} {\bibfnamefont {D.~R.}\ \bibnamefont {Amancio}},\
  }\bibfield  {title} {\enquote {\bibinfo {title} {Authorship attribution via
  network motifs identification},}\ }in\ \href@noop {} {\emph {\bibinfo
  {booktitle} {2016 5th Brazilian Conference on Intelligent Systems
  (BRACIS)}}}\ (\bibinfo {organization} {IEEE},\ \bibinfo {year} {2016})\ pp.\
  \bibinfo {pages} {355--360}\BibitemShut {NoStop}%
\bibitem [{\citenamefont {Alon}(2007)}]{alon2007network}%
  \BibitemOpen
  \bibfield  {author} {\bibinfo {author} {\bibfnamefont {U.}~\bibnamefont
  {Alon}},\ }\bibfield  {title} {\enquote {\bibinfo {title} {Network motifs:
  theory and experimental approaches},}\ }\href@noop {} {\bibfield  {journal}
  {\bibinfo  {journal} {Nature Reviews Genetics}\ }\textbf {\bibinfo {volume}
  {8}},\ \bibinfo {pages} {450} (\bibinfo {year} {2007})}\BibitemShut {NoStop}%
\bibitem [{\citenamefont {Lahav}\ \emph {et~al.}(2004)\citenamefont {Lahav},
  \citenamefont {Rosenfeld}, \citenamefont {Sigal}, \citenamefont
  {Geva-Zatorsky}, \citenamefont {Levine}, \citenamefont {Elowitz},\ and\
  \citenamefont {Alon}}]{lahav2004dynamics}%
  \BibitemOpen
  \bibfield  {author} {\bibinfo {author} {\bibfnamefont {G.}~\bibnamefont
  {Lahav}}, \bibinfo {author} {\bibfnamefont {N.}~\bibnamefont {Rosenfeld}},
  \bibinfo {author} {\bibfnamefont {A.}~\bibnamefont {Sigal}}, \bibinfo
  {author} {\bibfnamefont {N.}~\bibnamefont {Geva-Zatorsky}}, \bibinfo {author}
  {\bibfnamefont {A.~J.}\ \bibnamefont {Levine}}, \bibinfo {author}
  {\bibfnamefont {M.~B.}\ \bibnamefont {Elowitz}}, \ and\ \bibinfo {author}
  {\bibfnamefont {U.}~\bibnamefont {Alon}},\ }\bibfield  {title} {\enquote
  {\bibinfo {title} {Dynamics of the p53-mdm2 feedback loop in individual
  cells},}\ }\href@noop {} {\bibfield  {journal} {\bibinfo  {journal} {Nature
  Genetics}\ }\textbf {\bibinfo {volume} {36}},\ \bibinfo {pages} {147}
  (\bibinfo {year} {2004})}\BibitemShut {NoStop}%
\bibitem [{\citenamefont {Anastasiadou}, \citenamefont {Jacob},\ and\
  \citenamefont {Slack}(2018)}]{anastasiadou2018non}%
  \BibitemOpen
  \bibfield  {author} {\bibinfo {author} {\bibfnamefont {E.}~\bibnamefont
  {Anastasiadou}}, \bibinfo {author} {\bibfnamefont {L.~S.}\ \bibnamefont
  {Jacob}}, \ and\ \bibinfo {author} {\bibfnamefont {F.~J.}\ \bibnamefont
  {Slack}},\ }\bibfield  {title} {\enquote {\bibinfo {title} {Non-coding
  {R}{N}{A} networks in cancer},}\ }\href@noop {} {\bibfield  {journal}
  {\bibinfo  {journal} {Nature Reviews Cancer}\ }\textbf {\bibinfo {volume}
  {18}},\ \bibinfo {pages} {5} (\bibinfo {year} {2018})}\BibitemShut {NoStop}%
\bibitem [{\citenamefont {Shen-Orr}\ \emph {et~al.}(2002)\citenamefont
  {Shen-Orr}, \citenamefont {Milo}, \citenamefont {Mangan},\ and\ \citenamefont
  {Alon}}]{shen2002network}%
  \BibitemOpen
  \bibfield  {author} {\bibinfo {author} {\bibfnamefont {S.~S.}\ \bibnamefont
  {Shen-Orr}}, \bibinfo {author} {\bibfnamefont {R.}~\bibnamefont {Milo}},
  \bibinfo {author} {\bibfnamefont {S.}~\bibnamefont {Mangan}}, \ and\ \bibinfo
  {author} {\bibfnamefont {U.}~\bibnamefont {Alon}},\ }\bibfield  {title}
  {\enquote {\bibinfo {title} {Network motifs in the transcriptional regulation
  network of {E}scherichia coli},}\ }\href@noop {} {\bibfield  {journal}
  {\bibinfo  {journal} {Nature Genetics}\ }\textbf {\bibinfo {volume} {31}},\
  \bibinfo {pages} {64} (\bibinfo {year} {2002})}\BibitemShut {NoStop}%
\bibitem [{\citenamefont {Mangan}\ \emph {et~al.}(2006)\citenamefont {Mangan},
  \citenamefont {Itzkovitz}, \citenamefont {Zaslaver},\ and\ \citenamefont
  {Alon}}]{mangan2006incoherent}%
  \BibitemOpen
  \bibfield  {author} {\bibinfo {author} {\bibfnamefont {S.}~\bibnamefont
  {Mangan}}, \bibinfo {author} {\bibfnamefont {S.}~\bibnamefont {Itzkovitz}},
  \bibinfo {author} {\bibfnamefont {A.}~\bibnamefont {Zaslaver}}, \ and\
  \bibinfo {author} {\bibfnamefont {U.}~\bibnamefont {Alon}},\ }\bibfield
  {title} {\enquote {\bibinfo {title} {The incoherent feed-forward loop
  accelerates the response-time of the gal system of {E}scherichia coli},}\
  }\href@noop {} {\bibfield  {journal} {\bibinfo  {journal} {Journal of
  Molecular Biology}\ }\textbf {\bibinfo {volume} {356}},\ \bibinfo {pages}
  {1073--1081} (\bibinfo {year} {2006})}\BibitemShut {NoStop}%
\bibitem [{\citenamefont {Gorochowski}, \citenamefont {Grierson},\ and\
  \citenamefont {di~Bernardo}(2018)}]{gorochowski2018organization}%
  \BibitemOpen
  \bibfield  {author} {\bibinfo {author} {\bibfnamefont {T.~E.}\ \bibnamefont
  {Gorochowski}}, \bibinfo {author} {\bibfnamefont {C.~S.}\ \bibnamefont
  {Grierson}}, \ and\ \bibinfo {author} {\bibfnamefont {M.}~\bibnamefont
  {di~Bernardo}},\ }\bibfield  {title} {\enquote {\bibinfo {title}
  {Organization of feed-forward loop motifs reveals architectural principles in
  natural and engineered networks},}\ }\href@noop {} {\bibfield  {journal}
  {\bibinfo  {journal} {Science Advances}\ }\textbf {\bibinfo {volume} {4}},\
  \bibinfo {pages} {eaap9751} (\bibinfo {year} {2018})}\BibitemShut {NoStop}%
\bibitem [{\citenamefont {van Nes}, \citenamefont {Rip},\ and\ \citenamefont
  {Scheffer}(2007)}]{van2007theory}%
  \BibitemOpen
  \bibfield  {author} {\bibinfo {author} {\bibfnamefont {E.~H.}\ \bibnamefont
  {van Nes}}, \bibinfo {author} {\bibfnamefont {W.~J.}\ \bibnamefont {Rip}}, \
  and\ \bibinfo {author} {\bibfnamefont {M.}~\bibnamefont {Scheffer}},\
  }\bibfield  {title} {\enquote {\bibinfo {title} {A theory for cyclic shifts
  between alternative states in shallow lakes},}\ }\href@noop {} {\bibfield
  {journal} {\bibinfo  {journal} {Ecosystems}\ }\textbf {\bibinfo {volume}
  {10}},\ \bibinfo {pages} {17} (\bibinfo {year} {2007})}\BibitemShut {NoStop}%
\bibitem [{\citenamefont {Scheffer}\ and\ \citenamefont
  {Jeppesen}(2007)}]{scheffer2007regime}%
  \BibitemOpen
  \bibfield  {author} {\bibinfo {author} {\bibfnamefont {M.}~\bibnamefont
  {Scheffer}}\ and\ \bibinfo {author} {\bibfnamefont {E.}~\bibnamefont
  {Jeppesen}},\ }\bibfield  {title} {\enquote {\bibinfo {title} {Regime shifts
  in shallow lakes},}\ }\href@noop {} {\bibfield  {journal} {\bibinfo
  {journal} {Ecosystems}\ }\textbf {\bibinfo {volume} {10}},\ \bibinfo {pages}
  {1--3} (\bibinfo {year} {2007})}\BibitemShut {NoStop}%
\bibitem [{\citenamefont {Scheffer}\ \emph {et~al.}(2001)\citenamefont
  {Scheffer}, \citenamefont {Carpenter}, \citenamefont {Foley}, \citenamefont
  {Folke},\ and\ \citenamefont {Walker}}]{scheffer2001catastrophic}%
  \BibitemOpen
  \bibfield  {author} {\bibinfo {author} {\bibfnamefont {M.}~\bibnamefont
  {Scheffer}}, \bibinfo {author} {\bibfnamefont {S.}~\bibnamefont {Carpenter}},
  \bibinfo {author} {\bibfnamefont {J.~A.}\ \bibnamefont {Foley}}, \bibinfo
  {author} {\bibfnamefont {C.}~\bibnamefont {Folke}}, \ and\ \bibinfo {author}
  {\bibfnamefont {B.}~\bibnamefont {Walker}},\ }\bibfield  {title} {\enquote
  {\bibinfo {title} {Catastrophic shifts in ecosystems},}\ }\href@noop {}
  {\bibfield  {journal} {\bibinfo  {journal} {Nature}\ }\textbf {\bibinfo
  {volume} {413}},\ \bibinfo {pages} {591} (\bibinfo {year}
  {2001})}\BibitemShut {NoStop}%
\bibitem [{\citenamefont {Erd{\"o}s}\ and\ \citenamefont
  {R{\'e}nyi}(1959)}]{erdos1959random}%
  \BibitemOpen
  \bibfield  {author} {\bibinfo {author} {\bibfnamefont {P.}~\bibnamefont
  {Erd{\"o}s}}\ and\ \bibinfo {author} {\bibfnamefont {A.}~\bibnamefont
  {R{\'e}nyi}},\ }\bibfield  {title} {\enquote {\bibinfo {title} {On random
  graphs, i},}\ }\href@noop {} {\bibfield  {journal} {\bibinfo  {journal}
  {Publicationes Mathematicae (Debrecen)}\ }\textbf {\bibinfo {volume} {6}},\
  \bibinfo {pages} {290--297} (\bibinfo {year} {1959})}\BibitemShut {NoStop}%
\bibitem [{\citenamefont {Zemp}\ \emph {et~al.}(2017)\citenamefont {Zemp},
  \citenamefont {Schleussner}, \citenamefont {Barbosa}, \citenamefont {Hirota},
  \citenamefont {Montade}, \citenamefont {Sampaio}, \citenamefont {Staal},
  \citenamefont {Wang-Erlandsson},\ and\ \citenamefont
  {Rammig}}]{zemp2017self}%
  \BibitemOpen
  \bibfield  {author} {\bibinfo {author} {\bibfnamefont {D.~C.}\ \bibnamefont
  {Zemp}}, \bibinfo {author} {\bibfnamefont {C.-F.}\ \bibnamefont
  {Schleussner}}, \bibinfo {author} {\bibfnamefont {H.~M.}\ \bibnamefont
  {Barbosa}}, \bibinfo {author} {\bibfnamefont {M.}~\bibnamefont {Hirota}},
  \bibinfo {author} {\bibfnamefont {V.}~\bibnamefont {Montade}}, \bibinfo
  {author} {\bibfnamefont {G.}~\bibnamefont {Sampaio}}, \bibinfo {author}
  {\bibfnamefont {A.}~\bibnamefont {Staal}}, \bibinfo {author} {\bibfnamefont
  {L.}~\bibnamefont {Wang-Erlandsson}}, \ and\ \bibinfo {author} {\bibfnamefont
  {A.}~\bibnamefont {Rammig}},\ }\bibfield  {title} {\enquote {\bibinfo {title}
  {Self-amplified {A}mazon forest loss due to vegetation-atmosphere
  feedbacks},}\ }\href@noop {} {\bibfield  {journal} {\bibinfo  {journal}
  {Nature Communications}\ }\textbf {\bibinfo {volume} {8}},\ \bibinfo {pages}
  {14681} (\bibinfo {year} {2017})}\BibitemShut {NoStop}%
\bibitem [{\citenamefont {Hirota}\ \emph {et~al.}(2011)\citenamefont {Hirota},
  \citenamefont {Holmgren}, \citenamefont {van Nes},\ and\ \citenamefont
  {Scheffer}}]{hirota2011global}%
  \BibitemOpen
  \bibfield  {author} {\bibinfo {author} {\bibfnamefont {M.}~\bibnamefont
  {Hirota}}, \bibinfo {author} {\bibfnamefont {M.}~\bibnamefont {Holmgren}},
  \bibinfo {author} {\bibfnamefont {E.~H.}\ \bibnamefont {van Nes}}, \ and\
  \bibinfo {author} {\bibfnamefont {M.}~\bibnamefont {Scheffer}},\ }\bibfield
  {title} {\enquote {\bibinfo {title} {Global resilience of tropical forest and
  savanna to critical transitions},}\ }\href@noop {} {\bibfield  {journal}
  {\bibinfo  {journal} {Science}\ }\textbf {\bibinfo {volume} {334}},\ \bibinfo
  {pages} {232--235} (\bibinfo {year} {2011})}\BibitemShut {NoStop}%
\bibitem [{\citenamefont {Staal}\ \emph {et~al.}(2018)\citenamefont {Staal},
  \citenamefont {Tuinenburg}, \citenamefont {Bosmans}, \citenamefont
  {Holmgren}, \citenamefont {van Nes}, \citenamefont {Scheffer}, \citenamefont
  {Zemp},\ and\ \citenamefont {Dekker}}]{staal2018forest}%
  \BibitemOpen
  \bibfield  {author} {\bibinfo {author} {\bibfnamefont {A.}~\bibnamefont
  {Staal}}, \bibinfo {author} {\bibfnamefont {O.~A.}\ \bibnamefont
  {Tuinenburg}}, \bibinfo {author} {\bibfnamefont {J.~H.}\ \bibnamefont
  {Bosmans}}, \bibinfo {author} {\bibfnamefont {M.}~\bibnamefont {Holmgren}},
  \bibinfo {author} {\bibfnamefont {E.~H.}\ \bibnamefont {van Nes}}, \bibinfo
  {author} {\bibfnamefont {M.}~\bibnamefont {Scheffer}}, \bibinfo {author}
  {\bibfnamefont {D.~C.}\ \bibnamefont {Zemp}}, \ and\ \bibinfo {author}
  {\bibfnamefont {S.~C.}\ \bibnamefont {Dekker}},\ }\bibfield  {title}
  {\enquote {\bibinfo {title} {Forest-rainfall cascades buffer against drought
  across the {A}mazon},}\ }\href@noop {} {\bibfield  {journal} {\bibinfo
  {journal} {Nature Climate Change}\ }\textbf {\bibinfo {volume} {8}},\
  \bibinfo {pages} {539--543} (\bibinfo {year} {2018})}\BibitemShut {NoStop}%
\bibitem [{\citenamefont {Nobre}\ \emph {et~al.}(2016)\citenamefont {Nobre},
  \citenamefont {Sampaio}, \citenamefont {Borma}, \citenamefont
  {Castilla-Rubio}, \citenamefont {Silva},\ and\ \citenamefont
  {Cardoso}}]{nobre2016land}%
  \BibitemOpen
  \bibfield  {author} {\bibinfo {author} {\bibfnamefont {C.~A.}\ \bibnamefont
  {Nobre}}, \bibinfo {author} {\bibfnamefont {G.}~\bibnamefont {Sampaio}},
  \bibinfo {author} {\bibfnamefont {L.~S.}\ \bibnamefont {Borma}}, \bibinfo
  {author} {\bibfnamefont {J.~C.}\ \bibnamefont {Castilla-Rubio}}, \bibinfo
  {author} {\bibfnamefont {J.~S.}\ \bibnamefont {Silva}}, \ and\ \bibinfo
  {author} {\bibfnamefont {M.}~\bibnamefont {Cardoso}},\ }\bibfield  {title}
  {\enquote {\bibinfo {title} {Land-use and climate change risks in the
  {A}mazon and the need of a novel sustainable development paradigm},}\
  }\href@noop {} {\bibfield  {journal} {\bibinfo  {journal} {Proceedings of the
  National Academy of Sciences}\ }\textbf {\bibinfo {volume} {113}},\ \bibinfo
  {pages} {10759--10768} (\bibinfo {year} {2016})}\BibitemShut {NoStop}%
\bibitem [{\citenamefont {van Nes}\ \emph {et~al.}(2014)\citenamefont {van
  Nes}, \citenamefont {Hirota}, \citenamefont {Holmgren},\ and\ \citenamefont
  {Scheffer}}]{van2014tipping}%
  \BibitemOpen
  \bibfield  {author} {\bibinfo {author} {\bibfnamefont {E.~H.}\ \bibnamefont
  {van Nes}}, \bibinfo {author} {\bibfnamefont {M.}~\bibnamefont {Hirota}},
  \bibinfo {author} {\bibfnamefont {M.}~\bibnamefont {Holmgren}}, \ and\
  \bibinfo {author} {\bibfnamefont {M.}~\bibnamefont {Scheffer}},\ }\bibfield
  {title} {\enquote {\bibinfo {title} {Tipping points in tropical tree cover:
  linking theory to data},}\ }\href@noop {} {\bibfield  {journal} {\bibinfo
  {journal} {Global Change Biology}\ }\textbf {\bibinfo {volume} {20}},\
  \bibinfo {pages} {1016--1021} (\bibinfo {year} {2014})}\BibitemShut {NoStop}%
\bibitem [{\citenamefont {Staal}\ \emph {et~al.}(2015)\citenamefont {Staal},
  \citenamefont {Dekker}, \citenamefont {Hirota},\ and\ \citenamefont {van
  Nes}}]{staal2015synergistic}%
  \BibitemOpen
  \bibfield  {author} {\bibinfo {author} {\bibfnamefont {A.}~\bibnamefont
  {Staal}}, \bibinfo {author} {\bibfnamefont {S.~C.}\ \bibnamefont {Dekker}},
  \bibinfo {author} {\bibfnamefont {M.}~\bibnamefont {Hirota}}, \ and\ \bibinfo
  {author} {\bibfnamefont {E.~H.}\ \bibnamefont {van Nes}},\ }\bibfield
  {title} {\enquote {\bibinfo {title} {Synergistic effects of drought and
  deforestation on the resilience of the south-eastern {A}mazon rainforest},}\
  }\href@noop {} {\bibfield  {journal} {\bibinfo  {journal} {Ecological
  Complexity}\ }\textbf {\bibinfo {volume} {22}},\ \bibinfo {pages} {65--75}
  (\bibinfo {year} {2015})}\BibitemShut {NoStop}%
\bibitem [{\citenamefont {Zemp}\ \emph {et~al.}(2014)\citenamefont {Zemp},
  \citenamefont {Schleussner}, \citenamefont {Barbosa}, \citenamefont {Van~der
  Ent}, \citenamefont {Donges}, \citenamefont {Heinke}, \citenamefont
  {Sampaio},\ and\ \citenamefont {Rammig}}]{zemp2014importance}%
  \BibitemOpen
  \bibfield  {author} {\bibinfo {author} {\bibfnamefont {D.~C.}\ \bibnamefont
  {Zemp}}, \bibinfo {author} {\bibfnamefont {C.-F.}\ \bibnamefont
  {Schleussner}}, \bibinfo {author} {\bibfnamefont {H.}~\bibnamefont
  {Barbosa}}, \bibinfo {author} {\bibfnamefont {R.}~\bibnamefont {Van~der
  Ent}}, \bibinfo {author} {\bibfnamefont {J.~F.}\ \bibnamefont {Donges}},
  \bibinfo {author} {\bibfnamefont {J.}~\bibnamefont {Heinke}}, \bibinfo
  {author} {\bibfnamefont {G.}~\bibnamefont {Sampaio}}, \ and\ \bibinfo
  {author} {\bibfnamefont {A.}~\bibnamefont {Rammig}},\ }\bibfield  {title}
  {\enquote {\bibinfo {title} {On the importance of cascading moisture
  recycling in {S}outh {A}merica},}\ }\href@noop {} {\bibfield  {journal}
  {\bibinfo  {journal} {Atmospheric Chemistry and Physics}\ }\textbf {\bibinfo
  {volume} {14}},\ \bibinfo {pages} {13337--13359} (\bibinfo {year}
  {2014})}\BibitemShut {NoStop}%
\end{thebibliography}%

\end{document}


\maketitle 
\clearpage 
\newpage

\begin{figure}[htbp]
\centering
\includegraphics[width=.75\textwidth]{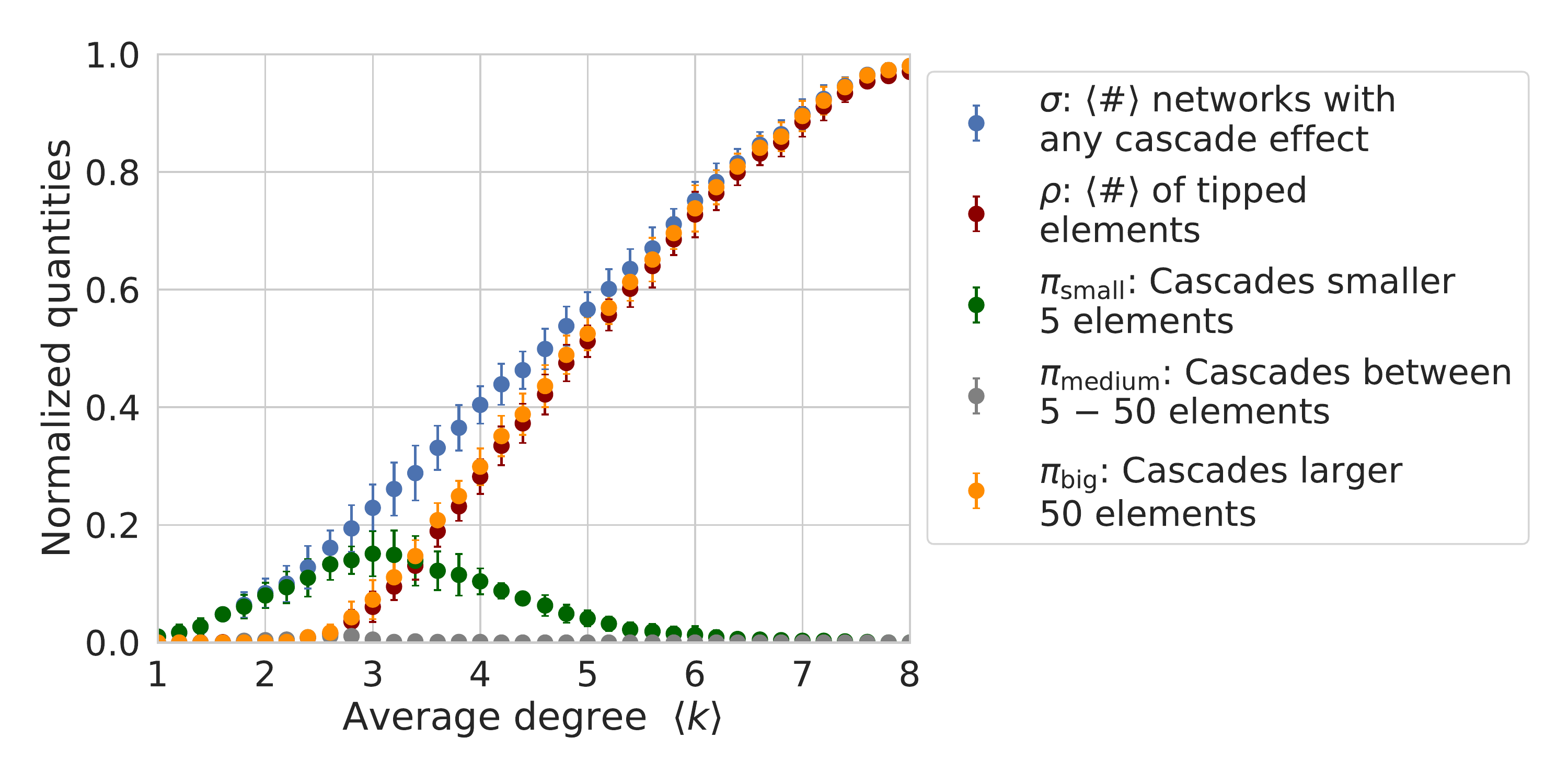}
\caption{Share of networks with any cascade effect $\sigma$ against average degree with a constant coupling constant of 0.162 (equal to the critical coupling constant of the feed forward loop). $\rho$ denotes the average number of tipped elements. Small $\pi_\text{small}$, medium $\pi_\text{medium}$ and big cascades $\pi_\text{big}$ cascade occurrences are shown, where small cascades involve at most 5 tipped nodes, medium cascades 5 $-$ 50 nodes and big cascades more than 50 tipped nodes. Each point was evaluated from 1000 Erd\H{o}s-Rényi networks with N = 100 and the standard deviation of occurrence are shown as error bars.}
\label{fig:supp:quantities}
\end{figure}

\begin{figure}[htbp]
\centering
\subfigure{\includegraphics[width=.61\textwidth]{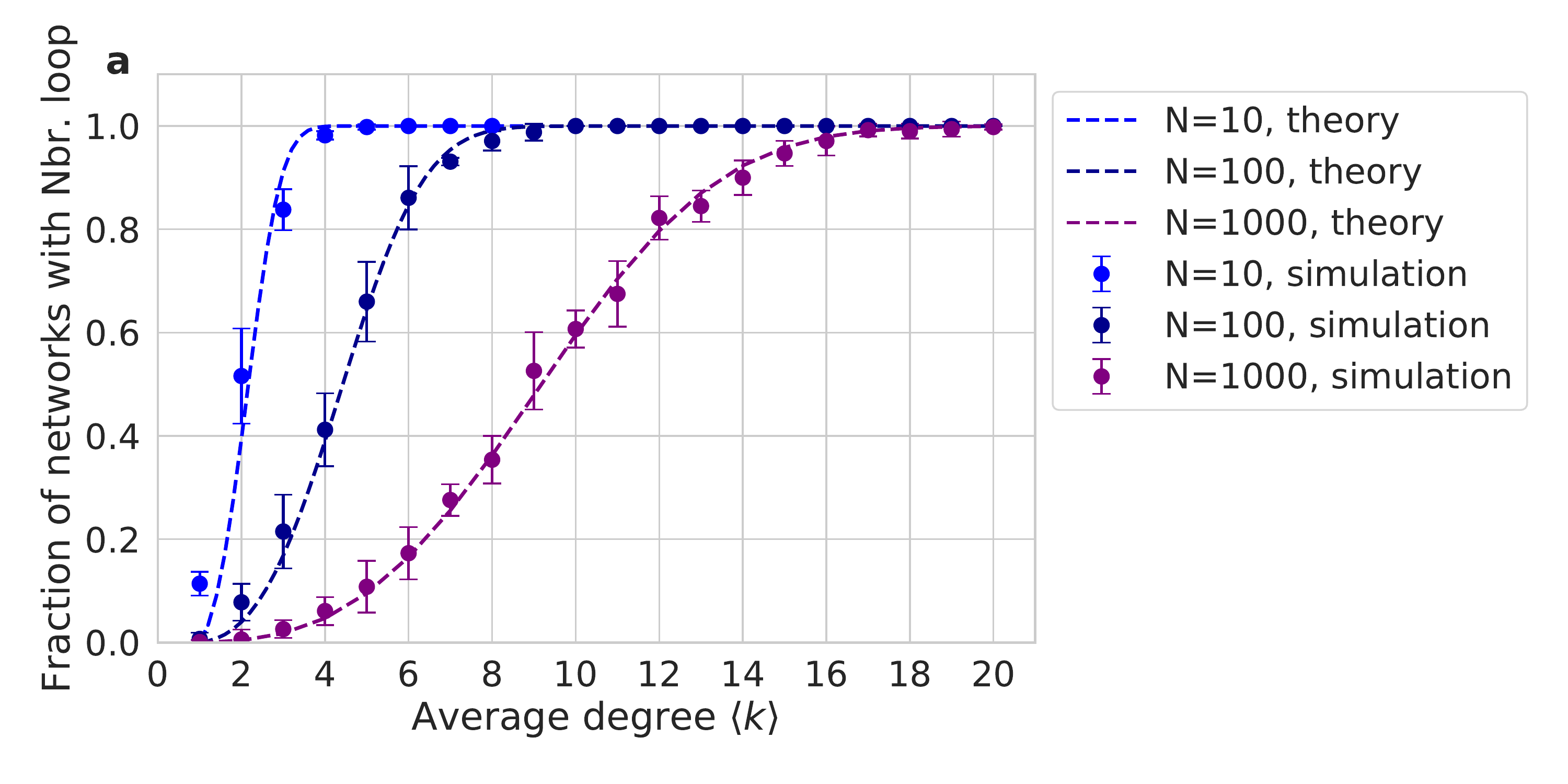}}
\subfigure{\includegraphics[width=.61\textwidth]{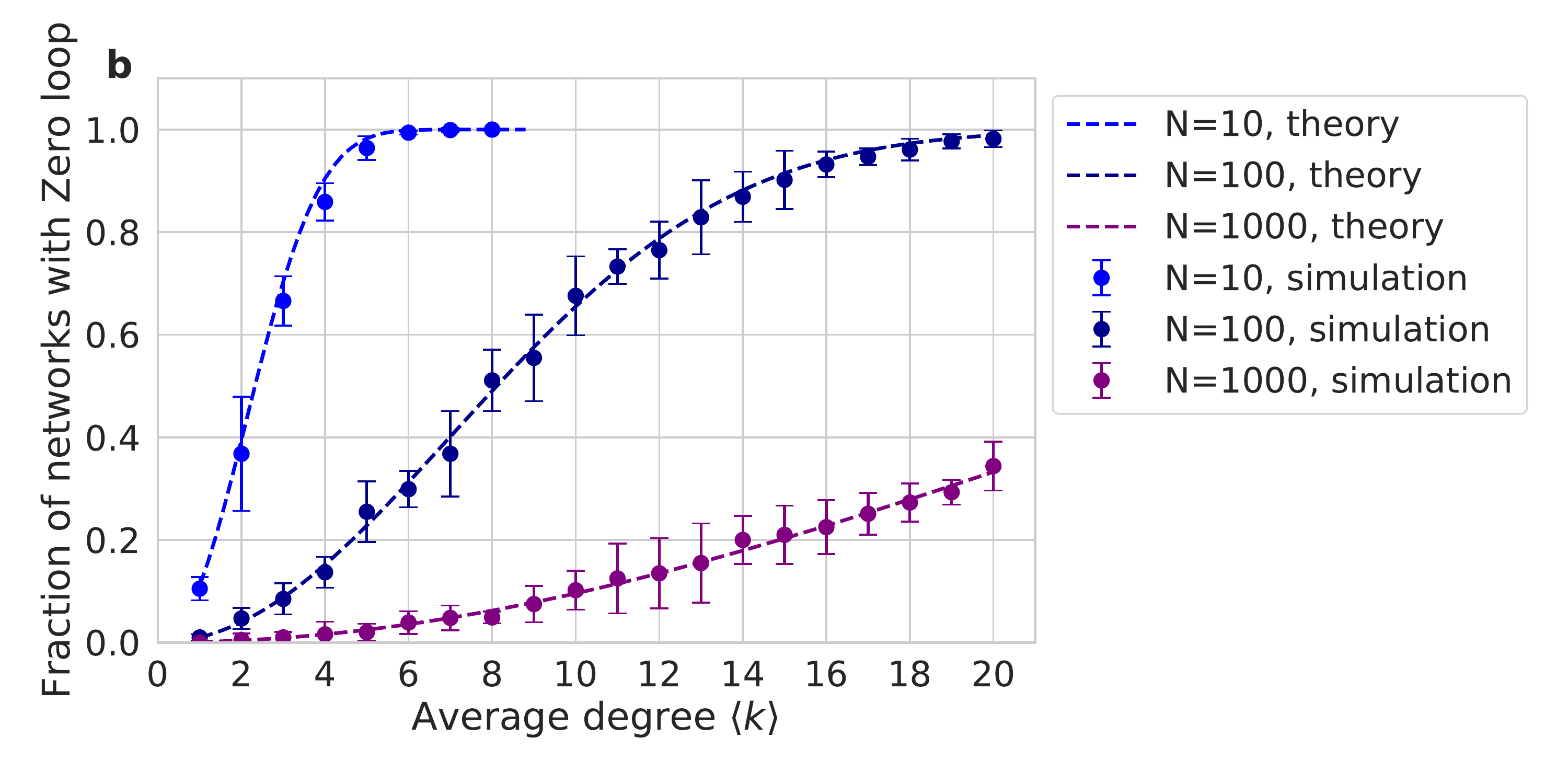}}
\subfigure{\includegraphics[width=.61\textwidth]{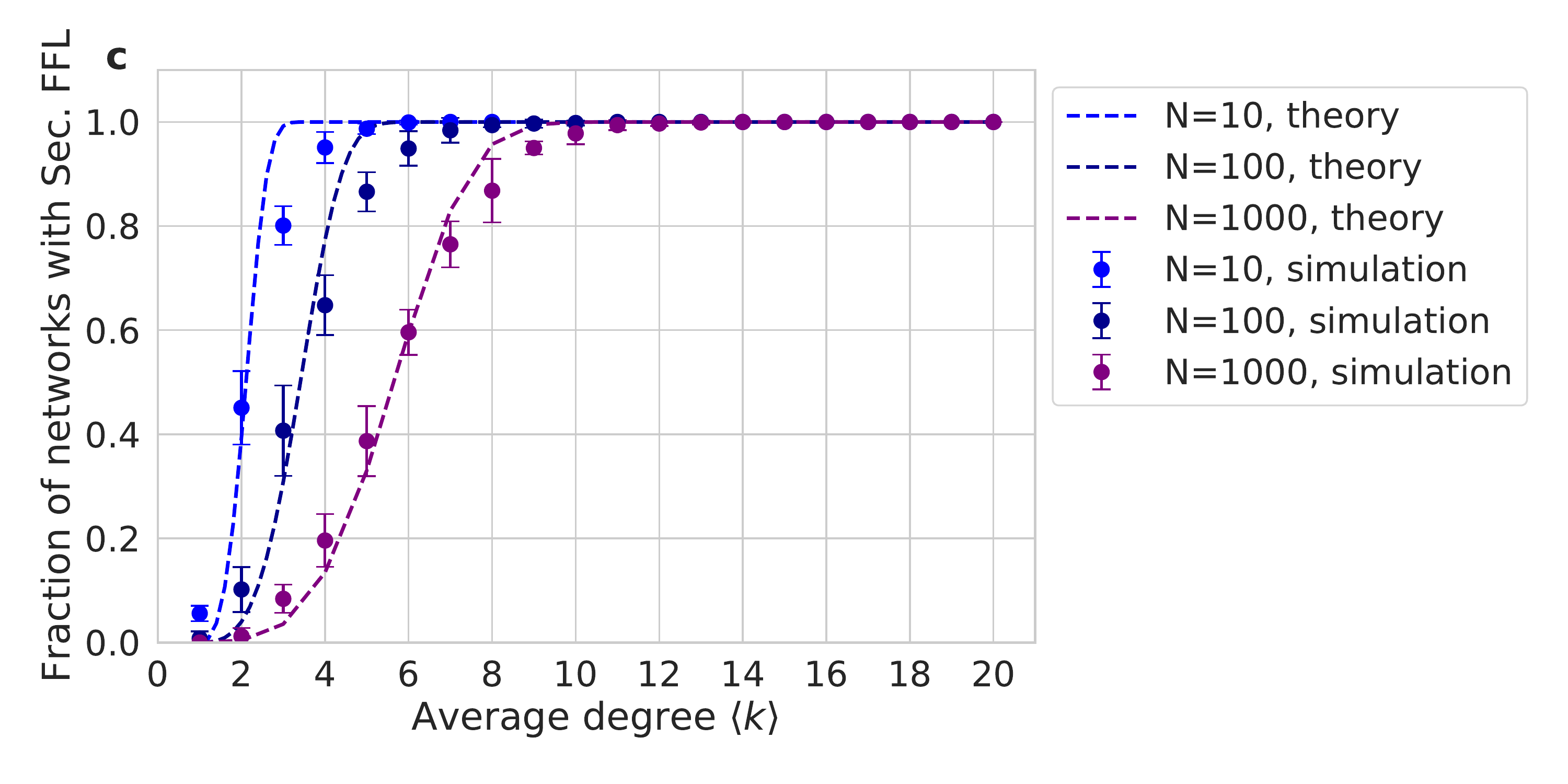}}
\caption{Scaling of the occurrence of the \textbf{a)} neighboring loop (theory from equation~5), \textbf{b)} zero loop (theory from equation~6) and \textbf{c)} the secondary feed forward loop (theory from equation~7) for networks of size 10, 100 and 1000. Error bars in both panels show the error in occurrence in 1000 realisations grouped as 10 $\times$ 100 samples.}
\label{fig:supp:scaling}
\end{figure}


\begin{figure}[htbp]
\centering
\includegraphics[width=\textwidth]{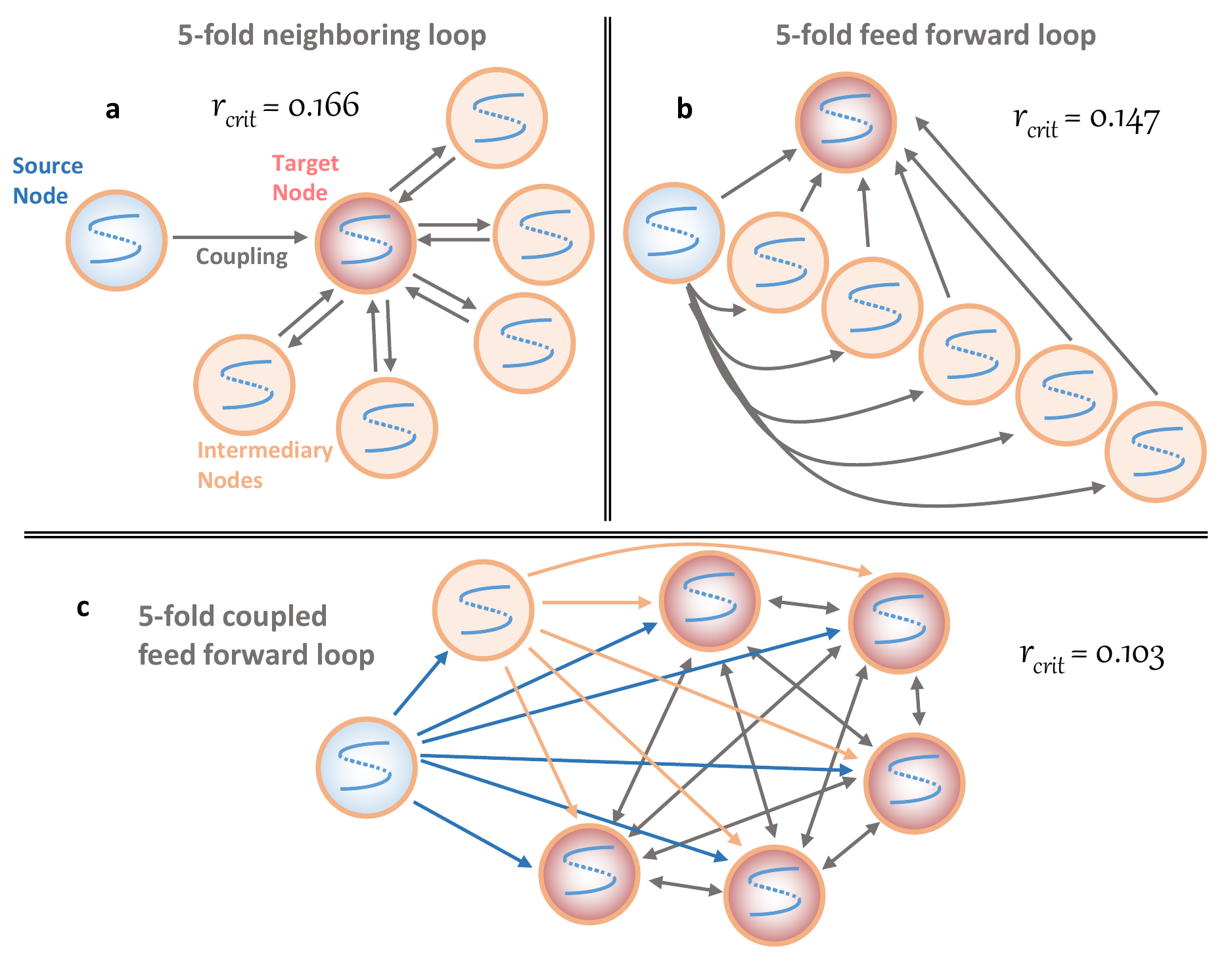}
\caption{Examples for five fold coupled motifs as evaluated in Fig.~4 in the main manuscript. \textbf{a)} 5-fold neighboring loop, \textbf{b)} 5-fold feed forward loop and \textbf{c)} 5-fold coupled feed forward loop. The target nodes are fully connected between each other. For better visibility, the couplings from the source node are blue and orange from the intermediary node and feedback loops are shown by a double headed arrow. All arrows have the same coupling strength. Furthermore, there are five target nodes since this motif is symmetric towards these nodes.}
\label{fig:supp:coupled_network_motifs}
\end{figure}


\begin{figure}[htbp]
\centering
\includegraphics[width=0.75\textwidth]{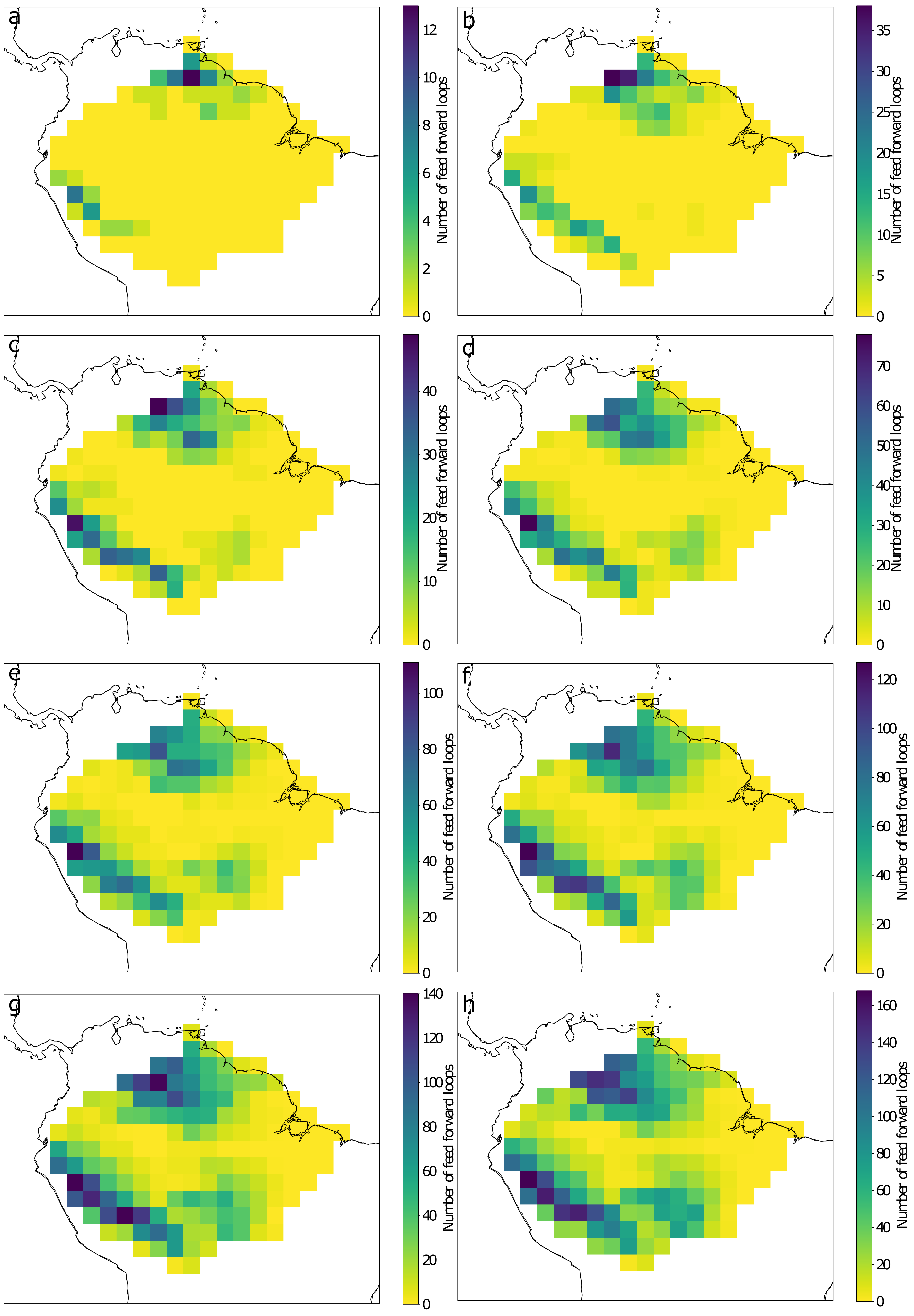}
\caption{Number of motifs that point to a certain cell in the 2$\times$2~$^\circ$ grid for the feed forward loop motif (similar for other motif types: zero loop, neighboring loop and secondary feed forward loop; not shown) for average degrees from \textbf{a) $-$ h)} 1 $-$ 8.}
\label{fig:supp:motif_ffl_degree_all}
\end{figure}\newpage\clearpage

\makeatletter
\renewcommand\@biblabel[1]{#1.}
\makeatother
\renewcommand\refname{Supplementary Bibliography}